\documentclass[a4paper,11pt]{article}
\pdfoutput=1 

\usepackage{jheppub} 

\usepackage{hyperref}
\usepackage[english]{babel}
\usepackage{amsmath,amssymb,bm,slashed,array}
\usepackage{amsfonts}
\usepackage{graphicx}
\usepackage{ifpdf}
\usepackage[sort&compress]{natbib}
\usepackage{color}
\usepackage[normalem]{ulem}
\usepackage{multirow}
\usepackage{booktabs}
\usepackage{subfigure}

\definecolor{nicered}{rgb}{0.7,0.1,0.1}
\definecolor{nicegreen}{rgb}{0.1,0.5,0.1}
\hypersetup{colorlinks,citecolor= nicegreen,linkcolor= nicered}
\definecolor{red}{rgb}{1.0, 0, 0}

\allowdisplaybreaks

\bibpunct{[}{]}{,}{n}{}{,}

\newcommand{\BR}{\mathcal B}

\providecommand{\abs}[1]{\lvert#1\rvert}

\newcommand{\ph}{\phantom}

\def\LjubljanaFMF{Faculty of Mathematics and Physics, University of Ljubljana,
                  Jadranska 19, 1000 Ljubljana, Slovenia }
\def\LjubljanaIJS{Jo\v zef Stefan Institute, Jamova 39, 1000 Ljubljana, Slovenia}
\def\MPHeidelberg{Max Planck Institut f\"ur Kernphysik, Saupfercheckweg 1, 69117 Heidelberg, Germany}

\title{Disentangling Flavor Violation in the Top--Higgs Sector at the LHC}

\author[a]{Admir Greljo,}
\author[a,b]{Jernej F.\ Kamenik}
\author[c]{and Joachim Kopp}

\affiliation[a]{\LjubljanaIJS}
\affiliation[b]{\LjubljanaFMF}
\affiliation[c]{\MPHeidelberg}

\emailAdd{admir.greljo@ijs.si}
\emailAdd{jernej.kamenik@ijs.si}
\emailAdd{jkopp@mpi-hd.mpg.de}

\abstract{ We study the LHC phenomenology of flavor changing Yukawa couplings between
  the top quark, the Higgs boson, and either an up or charm quark. Such $tuh$
  or $tch$ couplings arise for instance in models in which the Higgs sector is
  extended by the existence of additional Higgs bosons or by higher dimensional
  operators.  We emphasize the importance of anomalous single top plus Higgs
  production in these scenarios, in addition to the more widely studied $t \to
  h j$ decays.  By recasting existing CMS searches in multilepton and diphoton
  plus lepton final states, we show that bounds on $\BR(t\to hu)$ are improved
  by a factor of 1.5 when single top plus Higgs production is accounted for.
  We also recast the CMS search for vector boson plus Higgs production into
  new, competitive constraints on $tuh$ and $tch$ couplings, setting the limits
  of $\BR(t\to hu) < 0.7\%$ and $\BR(t\to hc) <1.2\%$.
  
  We then investigate the sensitivity of future searches in the multilepton
  channel and in the fully hadronic channel.  In multilepton searches, studying
  the lepton rapidity distributions and charge assignments can be used to
  discriminate between $tuh$ couplings, for which anomalous single top
  production is relevant, and $tch$ couplings, for which it is suppressed by
  the parton distribution function of the charm quark.  An analysis of fully
  hadronic $t+h$ production and $t\to h j$ decay can be competitive with the
  multilepton search at 100~fb$^{-1}$ of 13~TeV data if jet substructure
  techniques are employed to reconstruct boosted top quarks and Higgs bosons.
  To show this we develop a modified version of the HEPTopTagger algorithm,
  optimized for tagging $t \to h j$ decays.  Our sensitivity estimates on
  $\BR(t\to hu)$ ($\BR(t\to hc)$) at 100~fb$^{-1}$ of 13~TeV data for
  multilepton searches, vector boson plus Higgs search and fully hadronic
  search are $0.22\%$ ($0.33\%$), $0.15\%$ ($0.19\%$) and $0.36\%$ ($0.48\%$),
  respectively.}

\keywords{Higgs Physics, Top Physics, Beyond Standard Model}

\arxivnumber{1404.1278}

\begin{document}
\maketitle

\flushbottom

\section{Introduction}
\label{sec:intro}

Determining the properties of the newly-discovered Higgs boson is one of the
major goals of the LHC physics program. Higgs interactions with fermions are of
special interest since deviations from Standard Model (SM) predictions could
point to the existence of new flavor dynamics not too far above the electroweak
scale. Among the flavor violating Higgs couplings to quarks, the most
promising place to look for new physics at high energy colliders are processes
involving top quarks. On the one hand, all relevant indirect low energy
constraints on such processes are necessarily based on loop suppressed
observables~\cite{Harnik:2012pb}. On the other hand, the large number of top
quarks produced at the LHC allows us to study even strongly suppressed
contributions to top quark production and decay.
Using this feature, the CMS collaboration has provided the best official upper limit on
flavor violating $tch$ couplings: from a combination of
multilepton searches and diphoton plus lepton searches, 
the constraint $\BR(t\to hc) < 0.56\%$ is obtained at $95\%$ confidence level
(CL)~\cite{CMS:2014qxa}.

In the present work, we explore the LHC
sensitivity to non-standard flavor violating top--Higgs interactions ($tch$
and $tuh$) further.  Building upon related
theoretical~\cite{AguilarSaavedra:2000aj,Craig:2012vj,Wang:2012gp,Chen:2013qta,Atwood:2013ica,Agrawal:2013owa} and
experimental~\cite{CMS:2013jfa,TheATLAScollaboration:2013nia,Aad:2014dya} studies, we explore three
main directions: (1) We demonstrate the importance of the
single top+Higgs production processes in addition to $t \to h j$ decays. (2) We
demonstrate how these processes can be exploited to distinguish
$tch$ and $tuh$ couplings in leptonic $t+h$ events by studying
lepton rapidity distributions and charge assignments.
(3) we consider several novel search signatures including hadronic top
decays and Higgs decays to $b\bar b$ and $\tau^+\tau^-$. 
While this leads to more challenging signatures requiring efficient discrimination against the
large SM backgrounds, the final sensitivity is compensated by increased signal yields. 

The remainder of the paper is organized as follows: In Sec.~\ref{sec:setup} we set
up the notation and introduce our main physics ideas. Then we explore and
quantify these insights in more detail using several top and Higgs decay modes.
Multilepton searches~\cite{Craig:2012vj} are particularly sensitive to $(t\to b\ell
\nu) + (h \to W^+ W^-,\ ZZ,\ \tau^+\tau^-)$ final states, and in Sec.~\ref{sec:multi} we recast
a recent CMS analysis~\cite{CMS:2013jfa} to constrain these final states. 
In doing so, we demonstrate the importance of including the anomalous single top
production process $gu \to th$. In Sec.~\ref{sec:diphoton} we recast a recent CMS 
search~\cite{CMS:2014qxa} for flavor violating $tch$ coupling in the diphoton
plus lepton final state to set an improved bound on $tuh$ coupling.
In Sec.~\ref{sec:WH} we show that a competitive sensitivity can be obtained focusing specifically on  
$h\to \tau^+ \tau^-$ decays by recasting a CMS search~\cite{CMS:ckv} for associate $W+\text{Higgs}$ and
$Z+\text{Higgs}$ production.  We then proceed to future searches, showing in
Sec.~\ref{sec:multilepton-future} how a detailed analysis of kinematic distributions in
multilepton searches can be used to improve the sensitivity to both $tuh$ and $tch$ couplings, and to discriminate between them. 
Finally, in Sec.~\ref{sec:hadronic}, we develop a search strategy for the fully hadronic
final state $(t\to b \bar q q') + (h\to b\bar b)$, where for highly boosted
processes jet substructure techniques can be employed to identify top quarks
and Higgs bosons. 
We summarize our results in Sec.~\ref{sec:conclusions}.

\section{Flavor Violating Top--Higgs Couplings}
\label{sec:setup}

\begin{figure}
  \begin{center}
    \includegraphics[width=.6\textwidth]{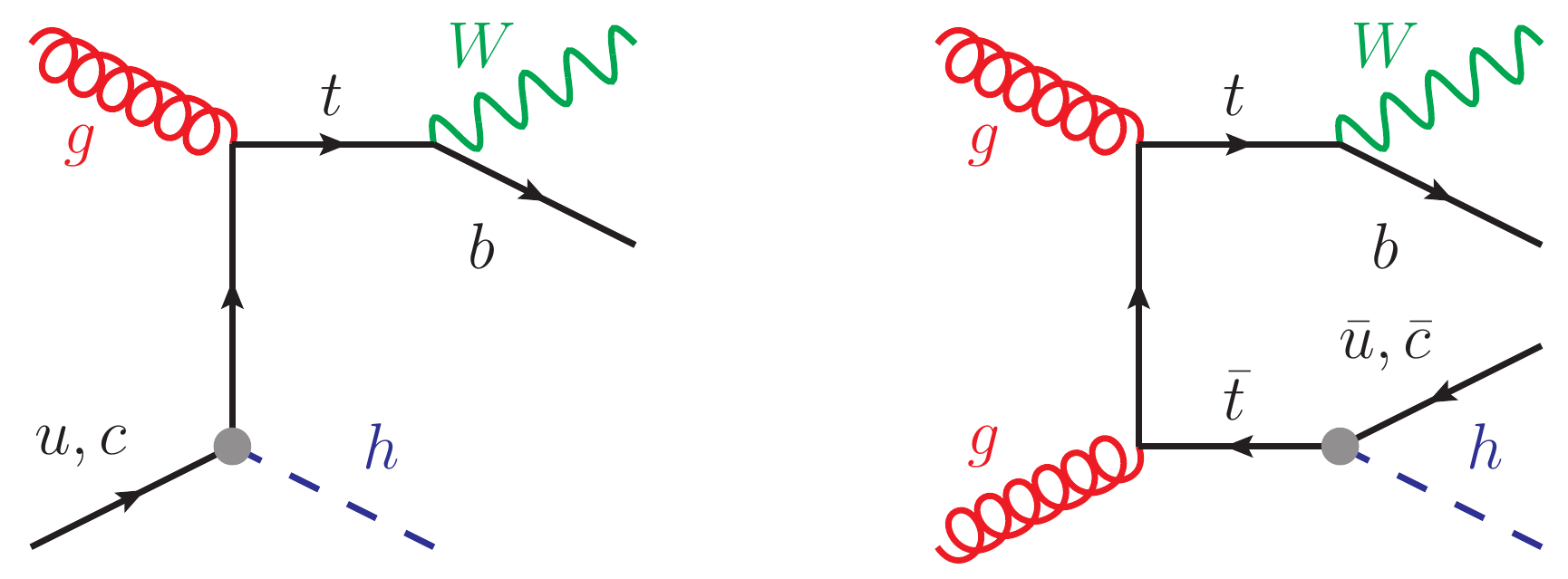}
  \end{center}
  \caption{\label{fig:diagrams} Example Feynman diagrams contributing to the
    LHC production of $pp\to (t\to W^+b) h$ (left) and $pp \to [(t\to W^+b)(\bar
    t \to h\bar q), (\bar t \to W^- \bar b) (t \to h q)]$ (right) through flavor
    violating top-Higgs interactions in Eq.~\eqref{lag:1} (marked with gray
    dots).}
\end{figure}

We parameterize the flavor violating top--Higgs interactions in the up-quark
mass eigenbasis as
\begin{equation}
  -\mathcal{L}_{tqh} = y_{tu}\,\bar{t}_{L}u_{R}h 
                   \,+\, y_{ut}\,\bar{u}_{L}t_{R}h
                   \,+\, y_{tc}\,\bar{t}_{L}c_{R}h
                   \,+\, y_{ct}\,\bar{c}_{L}t_{R}h
                   \,+\, \textrm{h.c.}\,.
  \label{lag:1}
\end{equation}
At tree level, this Lagrangian gives rise to the non-standard 3-body Higgs
boson decays $h\to t^*q \to W b q$ as well as the more interesting 2-body top
quark decays $t\to qh$, where $q=u,c$ (see Fig.~\ref{fig:diagrams}).
Neglecting the light quark masses and assuming the top quark decay width is
dominated by the SM value of $\Gamma(t\to Wb)$, the approximate relation
between the relevant $t\to qh$ branching ratios and the flavor violating Yukawa
couplings is given by
\begin{equation}
 \BR(t\to hq)
 = \frac{|y_{tq}|^2 + |y_{qt}|^2} {2\sqrt{2} G_F}
   \frac{(m_t^2 - m_h^2)^2}{(m_t^2 - m_W^2)^2 (m_t^2 + 2 m_W^2)}\eta_{QCD}
   \simeq 0.29 \big( |y_{tq}|^2 + |y_{qt}|^2 \big)\,,
\end{equation}
with the top quark mass $m_t$, the $W$ mass $m_W$, the Higgs mass $m_h$, and
the Fermi constant $G_F$.  The above expression is based on the leading order
formulae for both the $t \to W b$ and $t \to h q$ decay rates. The NLO QCD
correction to the branching ratio (in the pole top mass scheme) are included through
the factor $\eta_{QCD} = 1+0.97\alpha_s=1.10$, calculated using the known corrections to
the $t\to W^+ b$~\cite{Li:1990qf,Drobnak:2010wh} and $t\to c h$ 
decay widths~\cite{Zhang:2013xya}. We note that values of $y_{tq} =
y_{qt} \simeq 0.13$ correspond to $\BR(t\to hq) \simeq 1\%$.  Top quark
pair production followed by an anomalous $t \to qh$ decay has a total cross
section of
\begin{equation}
  \sigma[pp \to (th\bar q, \bar t h q)] = 2 \, \sigma(pp\to t\bar{t}) \, \BR(t\to hq)
    \simeq 140\ (470) \,\textrm{pb} \times \big( |y_{tq}|^2 + |y_{qt}|^2 \big) \,,
  \label{eq:sigma-tt}
\end{equation}
at the $\sqrt s = 8\ (13)$~TeV energy LHC, where we have used  the QCD NNLO
values of $\sigma(pp\to t\bar{t}) =  245\ (806)$~pb~\cite{Czakon:2013goa}.

\begin{figure}[!t]
  \begin{center}
    \includegraphics[width=.7\textwidth]{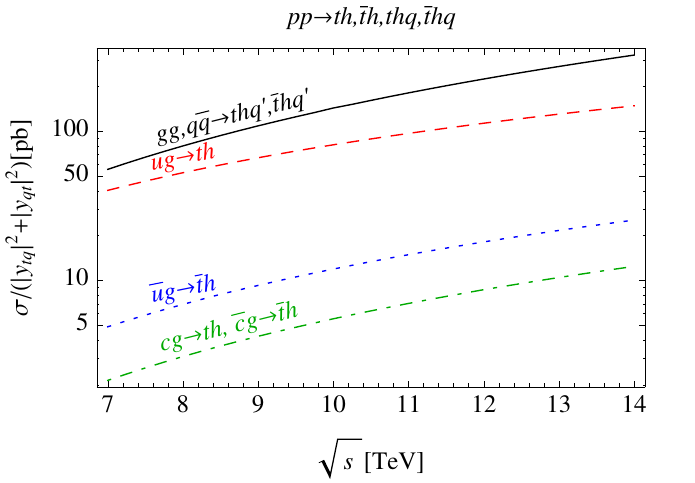}
  \end{center}
  \caption{\label{fig:xsections} Cross-sections for $(t \to b W) + (t \to h q)$
    and single top + Higgs production induced by flavor violating top-Higgs
    couplings as a function of the hadronic center of mass energy and
    normalized to the corresponding $tqh$ couplings. All partonic
    cross-sections are computed analytically at leading order in QCD, while
    parton luminosity integration is performed using MSTW2008 leading order
    parton distribution functions~\cite{Martin:2009iq}  with renormalization
    and factorization scales fixed to the top mass ($\mu_r = \mu_f = m_t =
    173.2$~GeV).}
\end{figure}

The interactions in Eq.~\ref{lag:1} also contribute to associated single top
plus Higgs production at the LHC. In particular the effects of $y_{tu}$ and
$y_{ut}$ are significant due to the large flux of valence $u$-quarks. The $t+h$
production cross-section is comparable in magnitude to \eqref{eq:sigma-tt}:
\begin{equation}
  \sigma(pp\to t h) \simeq 74\ (180)\ \text{pb} \times
                           \big( |y_{tu}|^2 + |y_{ut}|^2 \big) \,,
\end{equation}
where we have used the NLO QCD result of~\cite{Wang:2012gp,maltoni}. The cross
section for the conjugate process antitop + Higgs production is roughly an order
of magnitude smaller, and processes induced by $tch$ couplings are even more
suppressed as illustrated in Fig.~\ref{fig:xsections}. 
This implies that, for a given center of mass energy and luminosity, the
sensitivity to $tuh$  couplings is in general better than the one to
$tch$ couplings.

\begin{figure}
  \begin{center}
    \includegraphics[width=.6\textwidth]{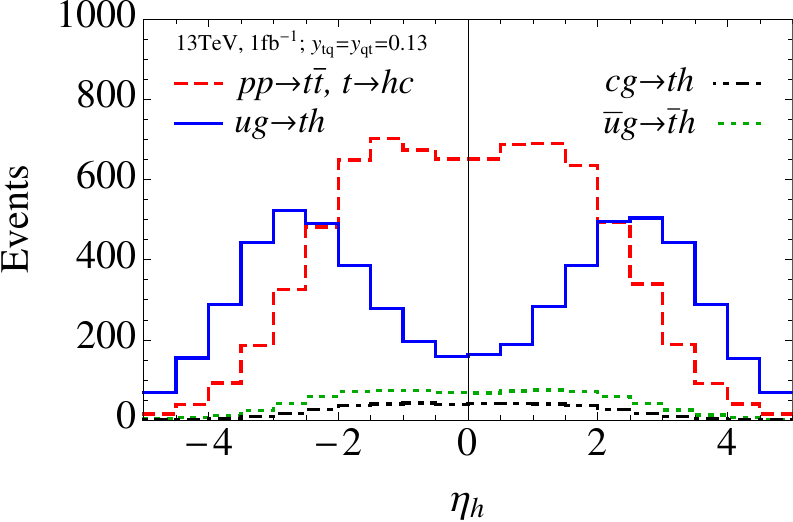}
  \end{center}
  \caption{Pseudorapidity distributions for the Higgs boson in various
    flavor violating processes at 13~TeV for $y_{tq} = y_{qt} = 0.13$
    (corresponding to $\BR(t\to hq) \simeq 1\%$) and an integrated
    luminosity of 1~fb$^{-1}$. The results are obtained using a
    FeynRules implementation of the effective interactions in
    Eq.~\eqref{lag:1} and using MadGraph for MC simulation. 
    Events are normalized to corresponding state of the art QCD corrected cross sections as discussed in the
    Sec.~\ref{sec:setup}.}
  \label{fig:higgs_eta}
\end{figure}

In addition, the presence or absence of a significant
contribution of $q g \to t h$ production in single top plus Higgs final states can be
used to distinguish between couplings to up quarks and couplings to charm
quarks.  A good discriminating variable is the Higgs boson pseudorapidity,
$\eta_h$, as illustrated in Fig.~\ref{fig:higgs_eta}.  
The relevance of this variable can be understood from the fact that in $u g$
scattering, the interaction products tend to be boosted in the direction of the
incoming valence $u$ quark, which on average carries a larger fraction of the
proton momentum than the gluon. In addition, the Higgs boson in such a
scattering process is preferentially produced in the direction of the up quark
in the partonic center of mass frame due to angular momentum conservation
combined with the quark chirality flip at the $tuh$ vertex. These effects add
up to make the resulting $\eta_h$ distribution peak at large rapidities.  For
initial states not containing valence quarks (gluon fusion-induced $t\bar t$
production as well as single top + Higgs production in $cg$, $\bar c g$, or
$\bar u g$ collision), both the top quark and Higgs boson are produced more
centrally.  Another useful handle on tagging single top plus Higgs production
in searches with leptonic top decays is the enhanced abundance of
positively charged leptons.

In the following sections we demonstrate the relevance of associated $t h$
production for probing flavor violating top--Higgs couplings using several
promising experimental signatures. Unless stated otherwise explicitly, all our numerical results are obtained using a
    FeynRules v1.6.16~\cite{Christensen:2008py} implementation of the effective interactions in
    Eq.~\eqref{lag:1} and using MadGraph~5, v1.5.11 (and v2.0.0-beta3)~\cite{Alwall:2011uj} for MC simulation. Furthermore we employ Pythia
v6.426~\cite{Sjostrand:2006za} for parton showering and hadronization, while Delphes~v3.0.9 (and v3.0.5)~\cite{deFavereau:2013fsa} is used
for detector simulation.

\section{Improved Limits on $tuh$ and $tch$ Couplings from Current LHC Searches}
\label{sec:improved-limits}

\subsection{Recasting the CMS Multilepton Search}
\label{sec:multi}

Multilepton searches at the LHC profit from relatively low SM backgrounds and
are therefore sensitive to new physics processes producing final states with many
leptons.  A good example is a final state with a top quark and a Higgs
boson~\cite{Craig:2012vj}, where the top quark decays to $b\ell\nu$, and the 126~GeV Higgs boson decays
to final states with up to four leptons.
The relevant processes are $h \to WW^* \to \ell\ell\nu\nu$,
$h \to \tau \tau$, $h \to ZZ^* \to \ell\ell jj$, $h \to ZZ^* \to \ell\ell\nu\nu$,
and $h\to ZZ^* \to \ell\ell\ell\ell$ with branching ratios $2.4\%$, $6.2\%$, $0.41\%$, $0.1\%$
and $0.03\%$, respectively~\cite{Heinemeyer:2013tqa}. Single top + Higgs production can 
thus yield up to five leptons, so that multilepton searches can be expected to constrain anomalous
flavor violating top--Higgs interactions.

In this section, we recast a recent CMS search for anomalous production of final states
with three or more isolated leptons~\cite{CMS:2013jfa}, based on $19.5~\textrm{fb}^{-1}$
of data at $\sqrt s = 8$~TeV. Data are binned into exclusive categories according
to the lepton flavor, the missing transverse energy $E_T^\text{miss}$, the
scalar sum of the transverse momenta of all the jets $H_T$, the existence of $b$-tagged
jets, and the presence or absence of opposite sign, same flavor (OSSF) light
lepton pairs. Events with an OSSF pair are further divided into ``below Z'',
``on Z'' and ``above Z'' categories based on the invariant mass of the OSSF
lepton pair relative to the $Z$ mass.

CMS has already interpreted this search as a constraint on the anomalous $tch$
coupling~\cite{CMS:2013jfa}, considering top pair production followed by
anomalous top decay to $h + j$.  However, the CMS search does not include contributions
from single top + Higgs production, which is irrelevant for $tch$ couplings, but
very important for $tuh$ couplings.  Therefore, we study in the following
the importance of associated $th$ production for constraining anomalous $tuh$
couplings.

We simulate the processes $pp \to t\bar{t}$ followed by $t \to hu$ or
$\bar{t}\to h\bar{u}$ decay, as well as $pp\to t h$ and $pp\to \bar{t} h$ using
MadGraph.  We rescale the
leading order cross sections to the corresponding higher order QCD
results. In particular, $pp \to t\bar{t}$ events are generated using the default MadGraph dynamical factorization 
and renormalization scales, and the final cross section is rescaled to $\sigma{(pp\to t\bar t)}=245$~pb~\cite{Czakon:2013goa}.
Single top plus Higgs events are generated using factorization 
and renormalization scales fixed to $\mu_f=\mu_r=m_h+m_t$, 
and a QCD correction factor of $K_{QCD}=1.5$ is applied~\cite{Wang:2012gp}. Higgs bosons and gauge bosons are
decayed using BRIDGE v2.24~\cite{Meade:2007js}, where the SM Higgs branching ratios are taken 
from~\cite{Heinemeyer:2013tqa}. Showering and hadronization are
simulated in Pythia, and Delphes is used
for detector simulation. We have modified the default
implementation of the CMS detector in Delphes by switching to the anti-$k_T$
jet algorithm with distance parameter $R=0.5$, by changing the light charged
lepton isolation criteria in accordance with~\cite{CMS:2013jfa}, and by
implementing the $b$ tagging efficiencies and mistag rates given
in~\cite{CMS:2013jfa} for the medium working point of the Combined Secondary
Vertex (CSV) algorithm.

We apply analysis cuts in accordance with those used in the CMS multilepton
search~\cite{CMS:2013jfa}. In particular, we require the leading charged lepton
in each event to have $p_T > 20$~GeV. Additional light charged leptons must
have $p_T > 10$~GeV, and all of them must be within $\abs{\eta} < 2.4$. Events
are rejected if they have an OSSF lepton pair with invariant mass $m_{\ell \ell} <
12$~GeV. Jets are required to have $\abs{\eta} < 2.5$ and $p_T > 30$~GeV, and
an angular distance $\Delta R > 0.3$ from any isolated charged lepton
candidates.

\begin{table}
\centering
  \begin{centering}
  \begin{tabular}{ccccccccc}
       & OSSF pair & $N_{\textrm{b-jets}}$ & $H_{T}(\textrm{GeV})$ & $E_{T}^{miss}(\textrm{GeV})$ & $N(t\to h j)$ & $N(th)$ & $N_{obs}$ & $N_{exp}$\tabularnewline
    \hline 
    1. & below Z & $\geq1$ & $\leq200$ & $50-100$ & 10.8 & 6.7 &  48 & $48\pm23$   \\
    2. & no OSSF & $\geq1$ & $\leq200$ & $50-100$ &  4.4 & 3.0 &  29 & $26\pm13$   \\
    3. & below Z & $\geq1$ & $\leq200$ & $\leq50$ &  6.8 & 3.8 &  34 & $42\pm11$   \\
    4. & no OSSF & $\geq1$ & $\leq200$ & $\leq50$ &  4.2 & 2.5 &  29 & $23\pm10$   \\
    5. & below Z & $\geq1$ & $>200$    & $50-100$ &  2.5 & 0.6 &  10 & $9.9\pm3.7$ \\
    6. & below Z & $\geq1$ & $>200$    & $\leq50$ &  2.0 & 0.4 &   5 & $10\pm2.5$  \\
    7. & below Z & 0       & $\leq200$ & $50-100$ &  9.2 & 5.1 & 142 & $125\pm27$  \\
    8. & no OSSF & 0       & $\leq200$ & $50-100$ &  4.0 & 2.5 &  35 & $38\pm15$   \\
    9. & above Z & $\geq1$ & $\leq200$ & $\leq50$ &  1.9 & 1.2 &  17 & $18\pm6.7$  \\
  \end{tabular}
  \end{centering}
  \caption{Number of signal events, expected background events and observed events
    in each event category of the CMS multilepton analysis~\cite{CMS:2013jfa} for
    $\mathcal{B}(t\to hu)=0.01$.  All bins contain exactly three isolated light
    charged leptons.\label{tab1}}
\end{table}

The results of our simulations are presented in Table~\ref{tab1}. The most sensitive
bins have exactly three isolated leptons and no hadronically decaying taus.
Signal predictions are given for $y_{ut} = y_{tu} = 0.13$ which corresponds to
$\mathcal{B}(t\to hu) = 0.01$. Taking into account the fact that we use a simplified detector
simulation, the predictions for top pair production $N(t\to h j)$, are in good
agreement with the results obtained by CMS~\cite{CMS:2013jfa}. This serves as
an important cross check of our simulation.

Table~\ref{tab1} confirms that for $tuh$ couplings the contribution of
associated $th$ production to the signal, $N(th)$, is of the same order as the
contribution from $t\bar{t}$ production followed by $t \to h j$ decay,
$N(t\to h j)$, as advocated before. Using the CL$_s$ method~\cite{Read:2002hq}, we derive the new 95\%~CL limits
\begin{align}
  \mathcal{B}(t\to hc) &< 1.5\% \,, \label{eq:multilepton-limit-thc} \\
  \mathcal{B}(t\to hu) &< 1.0\% \,. \label{eq:multilepton-limit-thu}
\end{align}
The corresponding limits on the flavor violating couplings are $\sqrt{|y_{tc}|^2 +
|y_{ct}|^2} < 0.227$ and $\sqrt{|y_{tu}|^2 + |y_{ut}|^2} < 0.186$. We have
checked that the minor difference between Eq.~\eqref{eq:multilepton-limit-thc}
and the CMS result $\mathcal{B}(t\to hc) <1.28\%$ is due to the contributions
of hadronic tau decays which we do not include in our analysis. Our main
conclusion, namely that the limit on $\mathcal{B}(t\to uh)$ is more stringent
than the limit on $\mathcal{B}(t\to ch)$ by a factor of 1.5 due to associated
$th$ production, is unaffected by this omission.

\subsection{Recasting the CMS Diphoton plus Lepton Search}
\label{sec:diphoton}

Recently, CMS has interpreted a search for extended Higgs sectors in the diphoton plus lepton 
final state~\cite{CMS:2013eua} as a constraint on flavor violating $tch$ coupling~\cite{CMS:2014qxa}, 
using $19.5~\textrm{fb}^{-1}$ of data collected at $\sqrt s = 8$~TeV. 
In the following, we use this search to constrain also $tuh$ couplings, taking into
account the contribution from associated top plus Higgs production.
\begin{table}
\centering
  \begin{tabular}{ccccccc}
       & $N_{\textrm{b-jets}}$ & $E_{T}^{miss}(\textrm{GeV})$ & $N(t\to h j)$ & $N(th)$ & $N_{obs}$ & $N_{exp}$\tabularnewline
    \hline 
    1.  & $\geq1$   & $50-100$  & 3.2 & 1.3 &  1 & $2.3\pm1.2$   \\
    2.  & $\geq1$   & $30-50$    &  2.2 & 0.92 &  2 & $1.1\pm0.6$   \\
    3. & $\geq1$    & $\leq30$   &  1.9 & 0.83 &  2 & $2.1\pm1.1$   \\
    4. &     0           & $50-100$  &  2.4 & 1.1 &  7 & $9.5\pm4.4$   \\
    5. & $\geq1$    & $>100$    &  0.82 & 0.49 &  0 & $0.5\pm0.4$ \\
    6. & 0               & $>100$    &  0.87 & 0.52 &   1 & $2.2\pm1.0$  \\
    7. & 0               & $30-50$   &  1.6 & 0.64 &  29 & $21\pm10$  \\
  \end{tabular}
  \caption{Number of signal events, expected background events and observed events
    in each event category of the CMS diphoton plus lepton analysis~\cite{CMS:2014qxa} for
    $\mathcal{B}(t\to hu)=0.01$.  All bins contain exactly one isolated light
    charged lepton and two isolated photons in the Higgs mass window.\label{tabdiphoton}}
\end{table}
We use MadGraph to simulate the signal processes induced by 
$tuh$ couplings, namely, top pair production followed by anomalous $t$ or $\bar t$ decay 
as well as associated single $t$ (and $\bar t$) plus Higgs production. 
Leptonic top decays as well as Higgs decays to pairs of photons are simulated using MadGraph 
where the implementation of the effective $h\gamma\gamma$ interaction is adopted from~\cite{heft}. 
The SM branching ratio for $h\to \gamma\gamma$ is taken to be
$0.23\%$~\cite{Heinemeyer:2013tqa}.  We rescale the leading order cross
sections to the corresponding higher order QCD corrected results as in
Sec.~\ref{sec:multi}. We simulate showering and hadronization effects in Pythia
and detector effects in Delphes.
We use the same implementation of the CMS detector in Delphes
as in Sec.~\ref{sec:multi}.

We closely follow the CMS search~\cite{CMS:2014qxa} in our analysis. In particular, 
we require one light charged lepton with $p_T > 10$~GeV and $\abs{\eta} < 2.4$. 
We require two photons with $p_T>40$~GeV ($p_T > 25$~GeV) for the leading 
(next to leading) photon and $\abs{\eta} < 2.5$. The diphoton invariant mass is 
required to be between $120$ and $130$~GeV.  Events are categorized into exclusive 
categories based on $E_T^{miss}$ and on the presence or absence of a bottom-tagged jet.

We summarize the results of our simulations in Table~\ref{tabdiphoton}.  The
most sensitive bins have a $b$-tagged jet and no hadronically decaying
taus~\cite{CMS:2014qxa}.  The predictions for signal yields are given for
$y_{ut} = y_{tu} = 0.13$ which corresponds to $\mathcal{B}(t\to hu) = 0.01$. We
validate our simulation by closely reproducing the predictions for top pair
production followed by anomalous top decay, $N(t\to h j)$, presented in Table~3
of~\cite{CMS:2014qxa}.  Finally, the contribution from associated $th$
production, $N(th)$, is competitive and thus important in the case of flavor
violating $tuh$ interactions. As before, we employ the CL$_s$
method~\cite{Read:2002hq} to derive the new 95\%~CL limits
\begin{equation}
  \mathcal{B}(t\to hc) < 0.66\% \; \; \; \textrm{and } \; \; \;
  \mathcal{B}(t\to hu) < 0.45\% , \label{eq:diphoton-limit-thq}
\end{equation}
where the corresponding limits on the flavor violating Yukawa couplings are
$\sqrt{|y_{tc}|^2 + |y_{ct}|^2} < 0.151$ and $\sqrt{|y_{tu}|^2 + |y_{ut}|^2} <
0.125$. The obtained limit on $tch$ couplings is in a good agreement with the
CMS result $\sqrt{|y_{tc}|^2 + |y_{ct}|^2} < 0.14$~\cite{CMS:2014qxa}. 

The search in the diphoton plus lepton final state sets the most competitive
current bounds on flavor violating $tqh$ interactions and will remain very
promising for future studies.  The current search is mainly limited by
statistics, so that further improvements are expected at larger integrated
luminosities. Improvements are also expected in the data-driven
background estimation by fitting the background shapes from the sidebands
around the Higgs mass window in the diphoton invariant mass~\cite{CMS:2013eua}.
We estimate the expected sensitivity to $\BR(t \to h q)$
at 100~fb$^{-1}$ (3000~fb$^{-1}$) and $\sqrt{s}=13$~TeV to improve by a factor
$\sim 4$ ($\sim 25$), based on naive scaling in cross section and luminosity.
Our rough estimate is in a good agreement with the dedicated study performed 
by the ATLAS collaboration~\cite{ATLAS:2013sensitivityHgamma}.

Furthermore, the advantage of this search with respect to other searches is an
explicit reconstruction of the Higgs boson which would be very useful in the
case of a positive signal.  Finally, as we will show in
Sec.~\ref{sec:multilepton-future}, the origin of the signal ($tuh$ or $tch$
couplings) could be disentangled by studying the Higgs pseudorapidity
distribution and the charges of the light charged lepton from the
top decay.

\subsection{Recasting the CMS Search for Vector Boson + Higgs Production}
\label{sec:WH}

In~\cite{CMS:ckv}, the CMS collaboration has searched for Higgs bosons produced
in association with a $W$ or $Z$ and decaying to $\tau^+\tau^-$.  This
final state is very similar to the one obtained from single top + Higgs
production, followed by $t \to W b$ and $h \to \tau^+\tau^-$, and from
$t\bar{t}$ production with one of the top quarks decaying to $(h \to
\tau^+\tau^-) + j$.  The CMS search can thus be recast to set limits on the
flavor changing $tuh$ and $tch$ couplings that we are interested in here.

In doing so, we consider only the $\ell\ell\tau_h$ final state consisting of
two light leptons (electrons or muons) and one hadronically decaying $\tau$.
This final state turns out to be more sensitive than $\ell\tau_h\tau_h$ (one
light lepton and two hadronic $\tau$'s) in the CMS search, and is therefore
also expected to give the best sensitivity in our case. In particular, the main
competing factors affecting the relative importance of the $\ell\ell\tau_h$ and
$\ell\tau_h\tau_h$ channels---the small leptonic branching ratio of the $\tau$
and the larger fake rate for hadronic $\tau$'s---affect the $h + W,\ Z$ channel in the
same way as our $t+h$ final state.  CMS also consider final states with four
light charged leptons, with at least two of them consistent with a $Z$ decay.
Since in the case of $t+h$ production or $t\bar{t}$ production followed by $t
\to h j$ decay, only events with the suppressed Higgs decay $h \to Z Z^*$ could
contribute to this final state, we do not consider it here.

We simulate the $t+h$ signal and the top and Higgs decays in MadGraph. Since in~\cite{CMS:ckv}, CMS have used 5.0~fb$^{-1}$
of data collected at $\sqrt{s} = 7$~TeV as well as 19.5~fb$^{-1}$ of data collected
at $\sqrt{s} = 8$~TeV, we simulate events for both center-of-mass energies. We rescale the leading order cross
sections to the corresponding higher order QCD corrected results as in
Sec.~\ref{sec:multi}.
We use TAUOLA~v2.5\cite{Jadach:1993hs} to decay the $\tau$ leptons and Pythia for parton showering and hadronization.  We choose
Delphes as a detector simulation, and we adapt the default
implementation of the CMS detector by adjusting the $p_T$-dependent $\tau$ tagging
efficiency and mistag rate to the values given in~\cite{Chatrchyan:2012zz} for the
loose working point of the HPS (``hadron plus strips'') algorithm.

In accordance with~\cite{CMS:ckv} we use the following cuts; we require exactly two
light leptons (electrons or muons), with the $p_T$ of the leading lepton larger
than 20~GeV and that of the subleading lepton larger than 10~GeV. Muons are
required to have a pseudorapidity $|\eta| < 2.4$, while for electrons the
requirement is $|\eta| < 2.5$. The leptons must have the same charge to
suppress $Z$ backgrounds, and the flavor combinations $\mu\mu$ and $e\mu$ are
allowed while $ee$ events are vetoed.  We also require one $\tau$-tagged jet
with $p_T > 20$~GeV and $|\eta| < 2.3$.  Extra jets are allowed, but events
containing a $b$-tagged jet with $p_T > 20$~GeV and $|\eta| < 2.4$ are vetoed to
suppress $t\bar{t}$ backgrounds.  Finally, the scalar sum of the lepton and
$\tau$ $p_T$'s is required to be larger than 80~GeV.

\begin{figure}
  \begin{tabular}{cc}
    \includegraphics[width=.48\textwidth]{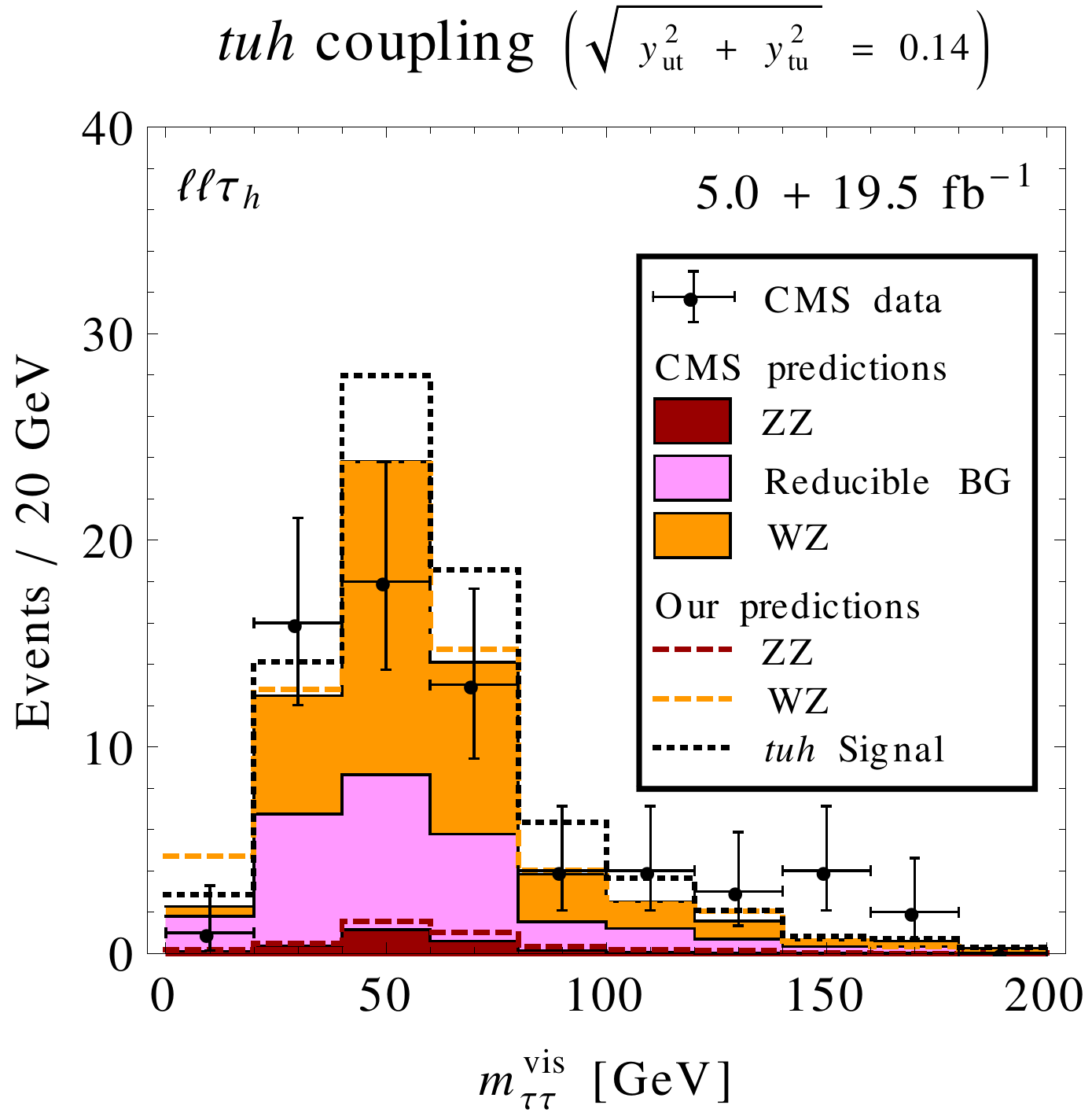} &
    \includegraphics[width=.48\textwidth]{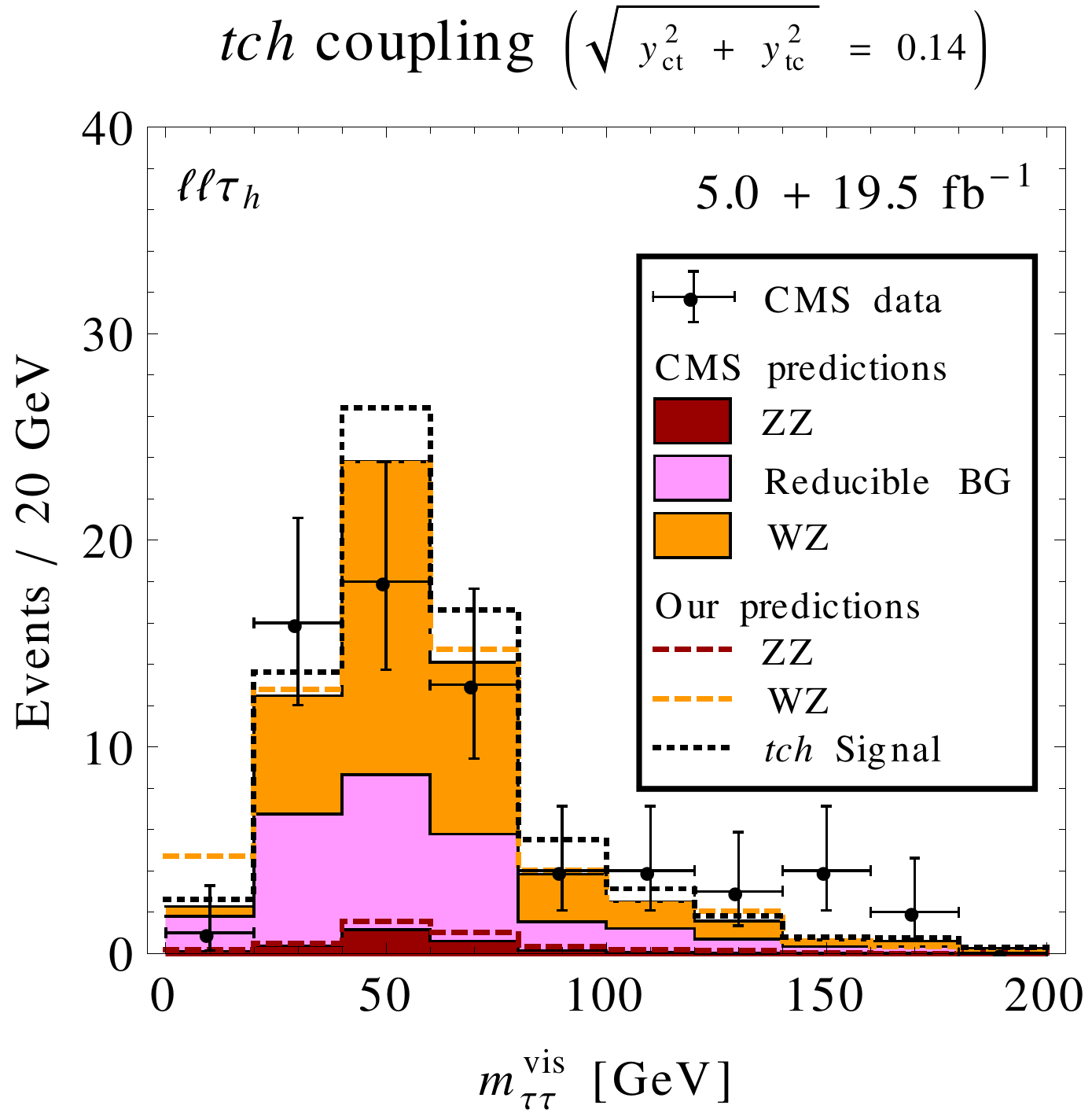} \\
  \end{tabular}
  \caption{Comparison of flavor violating $p p \to (t \to Wb) + (t \to h q)$
    and $p p \to t h$ signals to the data from a CMS search for vector boson + Higgs
    production~\cite{CMS:ckv} in the $\ell\ell\tau_h$ final state.  We plot the
    number of events against the invariant mass of the $\tau$ jet and the two
    light leptons, $m_{\tau\tau}^\text{vis}$.  Data points correspond to the
    CMS measurement in 5.0~fb$^{-1}$ of 7~TeV data and 19.5~fb$^{-1}$ of 8~TeV
    data.  The stacked shaded histograms show the CMS background prediction,
    which is in excellent agreement with our estimates of the $ZZ$ background
    (red dashed histogram) and the $WZ$ background (orange dashed histogram,
    stacked on top of the $ZZ$ and reducible backgrounds predicted by CMS).
    The black dotted histogram corresponds to the expected number of events
    (our signal prediction plus the CMS background prediction) in a model with
    flavor violating (a) top--up--Higgs couplings and (b) top--charm--Higgs
    couplings at the current upper limit $\sqrt{y_{qt}^2 + y_{tq}^2}
    = 0.14$ from CMS~\cite{CMS:2014qxa}.}
  \label{fig:vh-mtt-spectrum}
\end{figure}

\begin{table}
\centering
  \renewcommand{\arraystretch}{1.2}
  \begin{tabular}{lcccc}
      & $\sqrt{y_{ut}^2 + y_{tu}^2}$ & $\BR(t \to h u)$
      & $\sqrt{y_{ct}^2 + y_{tc}^2}$ & $\BR(t \to h c)$ \\ \hline
    Current limit
      & $< 0.16$  & $< 0.70 \times 10^{-2}$ & $< 0.21$  & $< 1.2  \times 10^{-2}$ \\
    Future sensitivity
      & $< 0.076$ & $< 0.15 \times 10^{-2}$ & $< 0.084$ & $< 0.19 \times 10^{-2}$
  \end{tabular}
  \caption{Limits on flavor changing $tuh$ and $tch$ couplings from recasting a CMS
    search for $V + (h \to \tau\tau)$ production~\cite{CMS:ckv} into a search for
    anomalous $t \to j h$ decays and anomalous single top + Higgs production using 5.0~fb$^{-1}$ of 7~TeV 
   and 19.5~fb$^{-1}$ of 8~TeV data. 
    We also show the expected sensitivity of a similar search using 100~fb$^{-1}$ of 13~TeV
    data.}
  \label{tab:vh-limits}
\end{table}

To verify our simulation and our analysis, we have also simulated the Standard
Model $ZZ$ and $WZ$ backgrounds.  Fig.~\ref{fig:vh-mtt-spectrum} shows that
our background predictions are in excellent agreement with the CMS
data~\cite{CMS:ckv} and with background predictions by CMS.  The figure also
shows that a $t+h$ signal induced by flavor violating top--Higgs couplings at
the current upper limit from CMS $\sqrt{y_{ut}^2 + y_{tu}^2} = 0.14$
would lead to a sizeable excess of events.  Quantifying this excess using the CL$_s$
method~\cite{Read:2002hq}, we find the new 95\%~CL limits on flavor-violating top
Yukawa couplings given in Table~\ref{tab:vh-limits}.

In the same table, we also give an estimate for the sensitivity of a future
$V+h$ search using 100~fb$^{-1}$ of 13~TeV LHC data and assuming identical cuts
as in the analysis at 7 and 8~TeV.  Since we cannot reliably model the
reducible background from fake leptons, we assume it to be of the same size and
have the same $m_{\tau\tau}^\text{vis}$ distribution as the $WZ$ background.
The larger instantaneous luminosity and larger pileup at 13~TeV may require
somewhat harder cuts and could lead to increased backgrounds from misidentified
jets.  We expect, however, that these complications can be offset by further
improvements of the analysis, for instance using multivariate techniques.

\section{Sensitivity of Future Searches}
\label{sec:future}

\subsection{Future Multilepton Searches and Discrimination between $tch$ and $tuh$ Couplings}
\label{sec:multilepton-future}

In this section, we study the potential of future multilepton searches at
13~TeV center of mass energy to constrain anomalous $tqh$ interactions or
to establish their existence.  Furthermore, we study the ability to
differentiate between $tuh$ and $tch$ couplings based on the
presence or absence of large contributions from associated single top plus
Higgs production to the signal.

\begin{figure}
  \begin{center}
    \includegraphics[width=.55\textwidth]{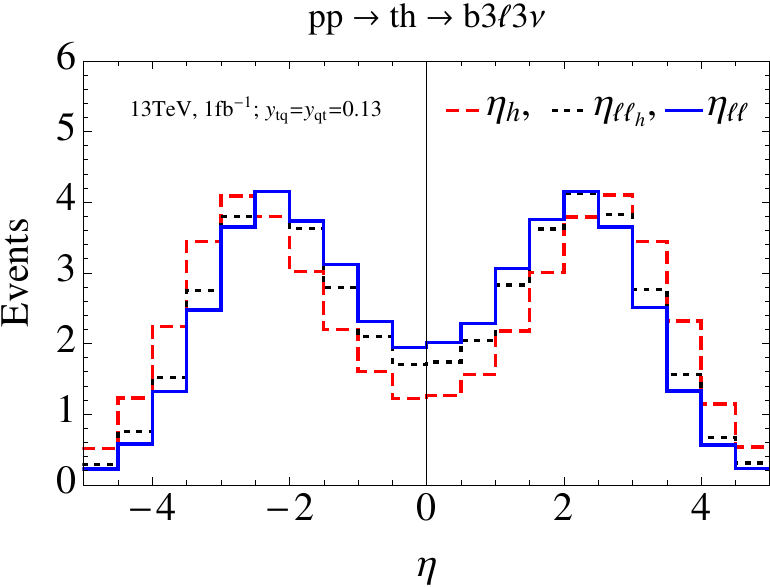}
  \end{center}
  \caption{For a parton level sample of $u g \to t h$ events with the decay chain $h\to WW^* \to \ell\ell\nu\nu$ 
    and $t\to Wb \to\ell\nu b$, we show the distributions of the Higgs
    pseudorapidity $\eta_h$ (red dashed), the pseudorapidity of the dilepton system
    from Higgs decay $\eta_{\ell\ell_h}$ (black dotted) and the pseudorapidity of the
    dilepton system $\eta_{\ell\ell}$ with the smallest angular distance
    $\Delta R_{\ell\ell}$ (blue solid).  Here we have assumed $y_{tu} = y_{ut} = 0.13$, a
    hadronic center of mass energy of 13~TeV and an integrated luminosity of
    1~fb$^{-1}$.}
  \label{fig:ugth_eta}
\end{figure}

We closely follow the analysis conducted in Sec.~\ref{sec:multi}. In
particular, we use the same lepton and jet reconstruction and isolation requirements as
before. An optimized search at 13~TeV will have 
slightly different requirements, such as somewhat higher lepton $p_T$ thresholds, but
we expect these to have only a minor impact on the sensitivity.
We require exactly three light charged leptons in the final state. In
order to differentiate between $tch$ and $tuh$ signals, we bin the data further
with respect to two variables: (1) the total sum of lepton charges $Q_{tot}$,\footnote{For related work see~\cite{Khatibi:2014via}.}
and (2) the pseudorapidity $\eta_{\ell\ell}$ of the opposite charge dilepton system with the
smallest angular distance $\Delta R_{\ell\ell} \equiv
\sqrt{\Delta\eta_{\ell\ell}^2+\Delta \phi_{\ell\ell}^2}$. We expect a $tuh$ signal
to have a preference for $Q_{tot} = +1$ due to a substantial contribution from
the process $ug \to th$, while $tch$ couplings
yield approximately equal numbers of events with $Q_{tot} = +1$ and $Q_{tot} =
-1$.  The idea behind the variable $\eta_{\ell\ell}$ is that the two leptons with
the smallest $\Delta R$ have the highest probability of originating from $h \to
WW^*$ decay (as opposed to a semileptonic top decay), so that $\eta_{\ell\ell}$
is an approximation to the pseudorapidity of the Higgs boson in the event, which
we have seen in Sec.~\ref{sec:setup} to be a promising discriminant between $tuh$ and $tch$
couplings. To illustrate the correlation between $\eta_{\ell\ell}$ and the Higgs rapidity
$\eta_h$, we have carried out a parton level simulation of the process $u g \to
t h$ followed by $h\to WW^* \to \ell\ell\nu\nu$ and $t\to Wb \to \ell\nu b$ using
MadGraph. In Fig.~\ref{fig:ugth_eta} we show
the resulting distributions for $\eta_h$, $\eta_{\ell\ell}$ and
$\eta_{\ell\ell_h}$. The latter quantity is defined as the rapidity of the
dilepton system that actually originates from Higgs decay.  We see that,
indeed, $\eta_{\ell\ell}$ nicely follows $\eta_h$.  Since we have already seen
in Sec.~\ref{sec:setup} and Fig.~\ref{fig:higgs_eta} that $\eta_h$ is an
efficient discriminator between $tuh$ and $tch$ couplings, we can expect the
same to hold for the experimentally accessible quantity $\eta_{\ell\ell}$.
We use two bins in $\eta_{\ell\ell}$: $|\eta_{\ell\ell}| > 1$ and $|\eta_{\ell\ell}| < 1$.

\begin{table}[!]
\centering
  \begin{centering}
\footnotesize{
    \begin{tabular}{cccc|cc|cc|cc}
      \multirow{3}{*}{} & \multirow{3}{*}{$H_{T}(\textrm{GeV})$}
        & \multirow{3}{*}{$E_{T}^{miss}(\textrm{GeV})$} & \multirow{3}{*}{$\left|\eta_{\ell \ell}\right|$}
        & \multicolumn{2}{c}{$N(\text{BG})$} & \multicolumn{2}{c|}{$N(t\to h j)$}
        & \multicolumn{2}{c}{$N(th)$} \\
      \cline{5-10} 
      &  &  &  & \multicolumn{2}{c|}{$Q_{tot}$} & \multicolumn{2}{c|}{$Q_{tot}$}
         & \multicolumn{2}{c}{$Q_{tot}$} \\
      &  &  &  & $-1$ & $+1$ & $-1$ & $+1$ & $-1$ & $+1$ \\
      \hline 
      \multirow{12}{*}{on OSSF} & \multirow{6}{*}{$\leq200$}
          & \multirow{2}{*}{$<50$}    & $<1$ &  61 & 67 & 20  & 19  & 2.5 & 7.4 \\
       &  &                           & $>1$ &  58 & 59 & 16  & 18  & 2.7 & 13  \\
       &  & \multirow{2}{*}{$50-100$} & $<1$ &  82 & 83 & 22  & 22  & 3.6 & 9.6 \\
       &  &                           & $>1$ &  77 & 88 & 20  & 21  & 2.9 & 16  \\
       &  & \multirow{2}{*}{$>100$}   & $<1$ &  34 & 32 & 7.0 & 5.7 & 1.2 & 3.7 \\
       &  &                           & $>1$ &  35 & 27 & 4.3 & 4.5 & 0.9 & 6.6 \\
      \cline{2-10}                                      
       & \multirow{6}{*}{$>200$}                        
          & \multirow{2}{*}{$<50$}    & $<1$ &  17 & 25 & 3.6 & 3.6 & 0.1 & 0.8 \\
       &  &                           & $>1$ &  19 & 21 & 2.3 & 2.1 & 0.2 & 1.3 \\
       &  & \multirow{2}{*}{$50-100$} & $<1$ &  35 & 30 & 4.7 & 5.3 & 0.2 & 0.8 \\
       &  &                           & $>1$ &  29 & 27 & 4.0 & 3.7 & 0.2 & 1.7 \\
       &  & \multirow{2}{*}{$>100$}   & $<1$ &  26 & 18 & 2.8 & 2.9 & 0.6 & 1.5 \\
       &  &                           & $>1$ &  21 & 18 & 1.8 & 1.8 & 0.2 & 2.8 \\
      \hline                                                                    
      \multirow{8}{*}{below Z} & \multirow{4}{*}{$\leq200$}          
          & \multirow{2}{*}{$50-100$} & $<1$ & 100 & 96 & 51  & 49  & 7.6 & 22  \\
       &  &                           & $>1$ &  83 & 93 & 42  & 42  & 7.3 & 34  \\
       &  & \multirow{2}{*}{$>100$}   & $<1$ &  36 & 42 & 12  & 15  & 1.8 & 8.6 \\
       &  &                           & $>1$ &  40 & 41 & 11  & 9.9 & 2.2 & 13  \\
      \cline{2-10}                                                              
       & \multirow{4}{*}{$>200$}                                                
          & \multirow{2}{*}{$50-100$} & $<1$ &  36 & 31 & 9.5 & 11  & 0.8 & 2.3 \\
       &  &                           & $>1$ &  23 & 20 & 7.8 & 10  & 0.6 & 3.7 \\
       &  & \multirow{2}{*}{$>100$}   & $<1$ &  22 & 20 & 8.1 & 7.7 & 0.6 & 3.1 \\
       &  &                           & $>1$ &  15 & 14 & 4.3 & 4.6 & 0.5 & 6.1 \\
      \hline                                                                    
      \multirow{12}{*}{above Z} & \multirow{6}{*}{$\leq200$}         
          & \multirow{2}{*}{$<50$}    & $<1$ &  42 & 39 & 7.8 & 7.9 & 1.3 & 3.1 \\
       &  &                           & $>1$ &  62 & 55 & 7.1 & 7.4 & 1.4 & 6.4 \\
       &  & \multirow{2}{*}{$50-100$} & $<1$ &  41 & 50 & 9.9 & 6.9 & 1.0 & 4.2 \\
       &  &                           & $>1$ &  68 & 71 & 8.2 & 8.8 & 1.2 & 7.9 \\
       &  & \multirow{2}{*}{$>100$}   & $<1$ &  20 & 21 & 2.1 & 2.3 & 0.5 & 2.6 \\
       &  &                           & $>1$ &  26 & 34 & 2.2 & 3.0 & 0.3 & 4.2 \\
      \cline{2-10}                                                              
       & \multirow{6}{*}{$>200$}                                                
          & \multirow{2}{*}{$<50$}    & $<1$ &  21 & 17 & 1.7 & 1.2 & 0.1 & 0.3 \\
       &  &                           & $>1$ &  29 & 27 & 1.7 & 1.9 & 0.0 & 0.9 \\
       &  & \multirow{2}{*}{$50-100$} & $<1$ &  22 & 28 & 2.4 & 2.6 & 0.2 & 0.3 \\
       &  &                           & $>1$ &  30 & 25 & 1.5 & 2.0 & 0.2 & 1.1 \\
       &  & \multirow{2}{*}{$>100$}   & $>1$ &  15 & 18 & 1.4 & 1.3 & 0.2 & 1.0 \\
       &  &                           & $<1$ &  22 & 20 & 1.7 & 0.7 & 0.1 & 2.1 \\
    \end{tabular}
}
  \end{centering}
  \caption{Number of predicted signal and background events per bin for a
    multilepton analysis using 100~fb$^{-1}$ of 13~TeV data. We have assumed
    $\mathcal{B}(t \to hu) = 0.01$.  The column labeled $N(t\to h j)$ shows the
    signal contribution from $t + (t \to h q)$ events, while the column labeled
    $N(th)$ contains the signal from single top plus Higgs production.  The
    column labeled $N(BG)$ is the expected background from SM $t\bar{t}$ events
    with a jet misidentified as a lepton.  For flavor violating $tch$ instead
    of $tuh$ couplings, $N(t\to h j)$ remains unchanged, while $N(th)$ becomes
    negligible because the process $g c \to t h$ compared to $g u \to t h$
    by the small parton distribution function for charm quarks.}
  \label{tab:sens}
\end{table}

Recalling the results of the analysis from Sec.~\ref{sec:multi} based on real
CMS data, we concentrate on the event categories that we have found to be most
sensitive: we consider only events with exactly three light charged leptons
that fall into the ``above Z'' , ``no OSSF'' or ``below Z'' categories; in the
latter case we also require $E^{miss}_T > 50$~GeV.  Moreover, we require at
least one $b$-tagged jet.  The dominant background in all categories is from
fully leptonic $t\bar{t}$ events with a jet misidentified as a
lepton~\cite{CMS:2013jfa}. We simulate $p p \to t \bar{t} \to
\ell^+\ell^-\nu\bar{\nu} b\bar{b}$ at 8~TeV and 13~TeV center of mass energy
using MadGraph and normalize the corresponding $pp\to t\bar{t}$ cross
sections to the NNLO QCD corrected values of $\sigma(pp\to t\bar t)=245\ (806)$~pb~\cite{Czakon:2013goa}, respectively.  Showering,
hadronization and detector effects are simulated using Pythia
 and Delphes. Following the procedure recommended by CMS~\cite{CMS:2013jfa}, we model fake leptons by randomly converting an isolated track to a
lepton with the measured conversion probability of 0.007 (0.006) for electron (muon) tracks.
To check the validity of this approach, we first compare
our 8~TeV predictions to CMS results~\cite{CMS:2013jfa} in the dilepton control
region that requires an opposite-sign $e\mu$ pair.
We obtain good agreement with the $H_T$ and $E_T^{miss}$ distributions shown in
Fig.~1 and Fig.~2 of \cite{CMS:2013jfa}. Second, we have checked that we agree
with CMS, at the level of 30--40\%, on the $E_T^{miss}$ distributions (provided
in~\cite{CMS:2013jfa}) of the $t\bar{t}$ background in the ``noOSSF'',
``above Z'' and ``below Z'' signal regions with low and high $H_T$ and with at
least one $b$-tagged jet. The
main difficulty in reproducing the background more precisely is the modeling of
lepton misidentification. Therefore, our quantitative results should be
considered with care, and a dedicated experimental analysis is clearly
necessary to obtain more precise predictions. We note in passing that the
irreducible SM background coming from associate top + Higgs production with a
cross-section of $\sigma_{th}^{SM} \simeq 74$~fb at 13 TeV LHC\footnote{This
value corresponds to the inclusive $pp \to t h j$ production cross section
calculated in MadGraph using 5-flavor
parton distribution functions and after applying a QCD correction factor of
$K_{QCD}=1.1$~\cite{Farina:2012xp}.} is only expected to become relevant once
the sensitivity reaches $\BR(t\to h q) \sim 10^{-4}$\,.

Our predicted signal and background yields at 13~TeV center of mass energy are
shown in Table~\ref{tab:sens} for an integrated luminosity of 100~fb$^{-1}$,
and for $\mathcal{B}(t \to hu) = 0.01$. The most sensitive bins fall into the
``below Z'' categories.  It is worth noting that single top + Higgs production
($N(th)$ column in Table~\ref{tab:sens}) tends to populate preferably bins with
$Q_{tot}=+1$ and $\left|\eta_{\ell \ell}\right|>1$, while the background
($N(BG)$ column) and the $t \to qh$ signal ($N(t\to h j)$ column) are much more
evenly distributed. This is of crucial importance in discriminating between the
$tuh$ and $tch$ signal hypotheses.

To estimate the achievable sensitivity and discovery reach for flavor violating
top--Higgs couplings, we use the CL$_s$ method~\cite{Read:2002hq}, treating all
bins as statistically independent Poisson variables.  We treat the overall
normalization of the background as a nuisance parameter (positively correlated
among all bins) to account for the uncertainty in our modeling of the lepton
misidentification probability. We do not impose any a priori constraints on the
nuisance parameter, i.e.\ we determine it in the analysis together with the
signal parameters, taking advantage of the fine-grained binning of the
simulated data.  Since in a realistic experimental analysis, the
misidentification rate can be measured from a $Z + \text{jets}$ control
sample~\cite{CMS:2013jfa}, our projected limits should be considered as very
conservative.

\begin{figure}
  \begin{center}
    \includegraphics[width=0.7\textwidth]{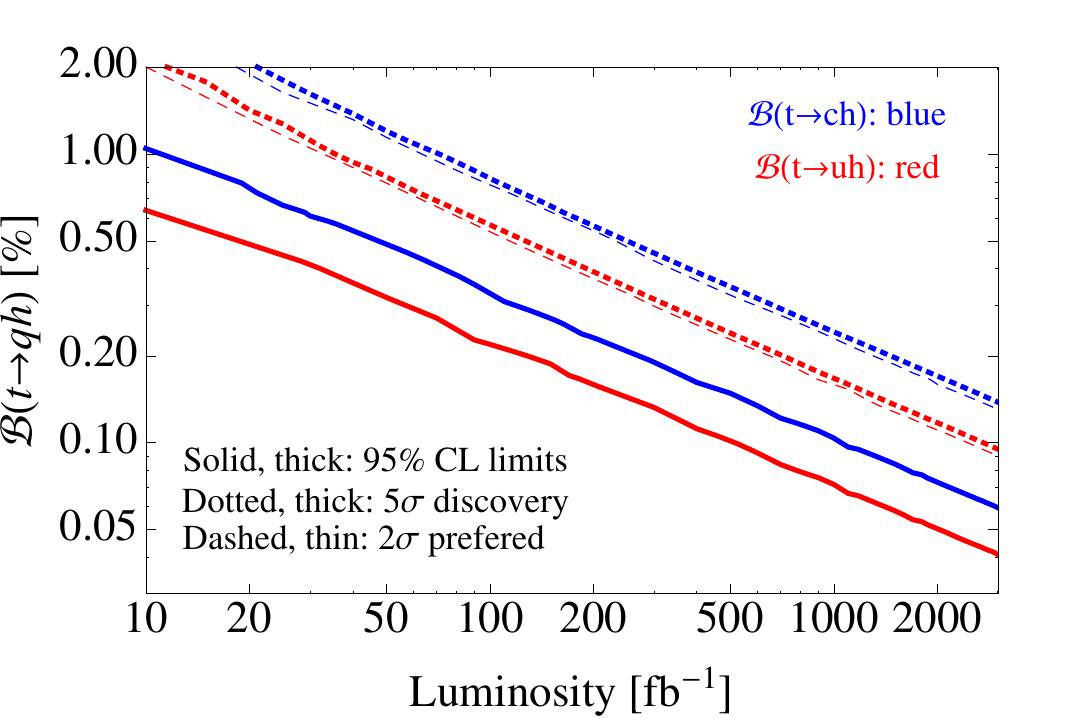}
  \end{center}
  \caption{Conservative estimates for the performance of an LHC search for flavor violating
    top--Higgs couplings in the multilepton channel at 13~TeV center of mass
    energy.  Thick solid lines represent the expected $95\%$~CL exclusion
    limits on $\BR (t \to h c)$ (blue) and $\BR (t \to h u)$
    (red) as a function of integrated luminosity. Thick dotted curves show the
    $5\sigma$ discovery potential.  For $tuh$ ($tch$) couplings above the thin
    dashed curves, the $tch$ ($tuh$) hypothesis can be excluded at 95\%~CL
    based on the different distributions of the dilepton rapidity
    $\eta_{\ell\ell}$ and the total charge $Q_{tot}$. The discrimination power of these
    variables comes from the presence or absence of the process $ug \to t h$.
    Since we treat the overall normalization of the background as an
    unconstrained nuisance parameter, our sensitivity projections are very
    conservative.}
  \label{fig:limits}
\end{figure}

The results of our statistical analysis are plotted in Fig.~\ref{fig:limits}.
Expected $95\%$~CL limits on $\BR (t\to hu)$ [$\BR (t\to hc)$] in
the absence of a signal are shown as red (blue) thick solid curves. The
expected $5\sigma$ discovery potential for a $tuh$ ($tch$) signal is shown as a
red (blue) thick dotted curve. The discrimination power between $tuh$ and $tch$
couplings is shown as thin dashed curves. For pure $tuh$ ($tch$) couplings
above the red (blue) thin dashed curve, the opposite hypothesis of pure $tch$
($tuh$) couplings can be ruled out at the 95\%~CL.

From the coincidence of the thick dotted curves and the thin dashed ones we
conclude that, if a $tqh$ signal is discovered (i.e.\ the BG only hypothesis is
rejected at $5\sigma$), the discrimination power between $tuh$ and $tch$
couplings is already at the level of $2\sigma$.  It is interesting that this
remarkable performance is achieved in spite of the rather generic, unoptimized
cuts in this multi-purpose multilepton analysis and of our rather conservative
treatment of systematic uncertainties.

\subsection{Searches in the Fully Hadronic Final State}
\label{sec:hadronic}

The final state with the largest branching ratio in $t+h$ production and
$t \to h q$ decay is the fully hadronic one. Modern jet substructure
techniques~\cite{Kaplan:2008ie, Plehn:2009rk, Plehn:2010st} 
offer promising tools to extract this signal from the otherwise overwhelming
background of QCD multijet events, SM $t \bar{t}$ and single top production and
vector boson plus jets production.  They are efficient when the
top quarks and Higgs bosons constituting the signal are highly boosted so that
the angular separation of their decay products is too small to be resolved
by conventional jet algorithms.  Instead, jet substructure methods use
``fat jets'', i.e.\ jets with a very large radius $R$.  After the initial
clustering, the fat jet is partially unclustered again to examine the invariant
mass of its largest subclusters. Comparing these invariant masses to the masses
of possible parent particles such as top quarks, Higgs bosons or $W$ bosons, the
algorithm decides how probable it is that the fat jet was produced by one of
these parent particles.

Here, we study the sensitivity of two analyses using jet substructure: (1) a
search for $t\bar{t}$ events with one SM top decay and one flavor
violating decay $t \to j + (h \to b \bar{b})$; (2) a search for anomalous
single top + Higgs production with SM top decay and $h \to b \bar{b}$.

In both analyses, we use the Cambridge-Aachen
algorithm~\cite{Dokshitzer:1997in} as implemented in
FastJet~3.0.3~\cite{Cacciari:2011ma} to cluster fat jets with a radius $R =
1.5$ and a minimum transverse momentum $p_T > 170$~GeV. We run
HEPTopTagger~v1.0~\cite{Plehn:2009rk, Plehn:2010st} with default settings on
these jets to identify those which are most likely to originate from a SM
hadronic top decay $t \to b + (W \to j j)$. HEPTopTagger imposes cuts on the
invariant masses of the three main subjets of the top candidate, requiring that
two of them reconstruct to a $W$, while all three together yield the top mass.
Moreover, their combined $p_T$ has to exceed 200~GeV.  In addition to these
kinematic cuts, we also require the subjet that is most likely to originate
from the $b$ quark to contain a $b$ tag (see Appendix~\ref{sec:th-tagger} for
details on our implementation of $b$-tagging).

\subsubsection{Analysis~1: $th$ tag + top tag}

To identify flavor violating decays $t \to j + (h \to b \bar{b})$ for analysis
1) and assign a ``$th$'' tag to the corresponding fat jets, we reprocess all
fat jets using a modified version of HEPTopTagger, which we have optimized for
this non-standard decay mode (see Appendix~\ref{sec:th-tagger}).  We require a
$b$ tag in each of the two subjets most likely to originate from the Higgs
decay.  We consider two different working points for our $th$ tagger: a loose
one with very robust kinematic cuts on the subjet invariant masses, and a tight
one with somewhat more restrictive cuts that make it more efficient at
suppressing backgrounds, but also more prone to systematic uncertainties in our
simulations. Details on the kinematic cuts are given in
Appendix~\ref{sec:th-tagger}.  A tight $th$ tag moreover requires that the fat
jet does not simultaneously carry a regular top tag.

Event selection for analysis~1 requires one fat jet with a loose or tight $th$
tag and a second fat jet with a top tag.

We consider the backgrounds from $t \bar{t}$ production, single
top production and QCD multijet production, but we have checked that $W +
\text{jets}$, $Z + \text{jets}$, $t \bar{t} + h$ and SM single top + Higgs
contributions are several orders of magnitude smaller than these dominant
backgrounds.  To simulate the $t \bar{t}$ and multijet backgrounds, we use
Sherpa~1.4.3~\cite{Gleisberg:2008ta, Schumann:2007mg, Gleisberg:2008fv,
Hoeche:2009rj, Schonherr:2008av} at leading order.  For $t \bar{t}$, we rescale
the cross section to the NNLO value $\sigma(pp\to t\bar t)=806$~pb~\cite{Czakon:2013goa}, while QCD
multijet events are rescaled by a factor $K=1.05$, which has been empirically found
to bring Sherpa predictions into agreement with data~\cite{Aad:2011tqa}.
We note that in a realistic experimental analysis, backgrounds
could be estimated directly from data.  For $t \bar{t}$ and single top events,
semileptonic final states offer a good control sample, while for QCD jet
production, anti-$b$-tags can be employed to define a control region.  For the
simulation of the SM single top background and of the signal we use MadGraph, followed by Pythia
 for parton showering and hadronization.

The predicted event counts after cuts from analysis~1 are shown in the upper part
of Table~\ref{tab:hadronic-event-rates} for $\sqrt{y_{qt}^2 + y_{tq}^2} = 0.1$
and assuming 100~fb$^{-1}$ of 13~TeV data. The predicted CL$_s$ sensitivity of the
analysis is summarized in the upper part of Table~\ref{tab:hadronic-sensitivity}.
We see that an analysis of the fully hadronic final state can improve upon the
current limits on flavor violating Higgs couplings, and that the future sensitivity
is only slightly worse than the one expected from analyses involving leptons.
A combined analysis of leptonic and hadronic final states would therefore seem
worthwhile. Moreover, the hadronic channel would provide a crucial cross-check in
case a signal is discovered in one of the other searches.

\begin{table}
\centering
  \begin{minipage}{14cm}
    \renewcommand{\arraystretch}{1.2}
    \begin{tabular}{p{3cm}rrr|rr|rr}
      & \multicolumn{3}{c|}{Background}
        & \multicolumn{2}{c|}{$\sqrt{y_{ut}^2 + y_{tu}^2} = 0.1$}
        & \multicolumn{2}{c}{$\sqrt{y_{ct}^2 + y_{tc}^2} = 0.1$} \\
      & $t\bar{t}$  & single-$t$ & QCD & $t \to h u$ & $t + h$ & $t \to h c$ & $t + h$ \\ \hline

      \multicolumn{4}{l|}{\bf Analysis~1: $th$ tag + top tag} & & & & \\
        \quad loose $th$ tags & 3\,510&   5.5\ph{0}&   125& 70\ph{.0}&  4.0     & 69\ph{.0}&0.57 \\

        \quad tight $th$ tags &    324&   0.52     &    85& 28\ph{.0}&  1.1     & 26\ph{.0}&0.15 \\ \hline

      \multicolumn{4}{l|}{\bf Analysis~2: Higgs tag + top tag} & & & & \\
        \quad preselection    &14\,800& 113\ph{.00}&4\,125&152\ph{.0}&120\ph{.0}&209\ph{.0}&14.0\ph{0} \\
        \quad final cuts      &    450&   2.3\ph{0}&    71&  6.9     & 32.6     &  8.4     & 1.1\ph{0} \\
    \end{tabular}
  \end{minipage}
  \caption{Predicted signal and background event rates in 100~fb$^{-1}$
    of 13~TeV data for the different variants of our fully hadronic analysis.}
  \label{tab:hadronic-event-rates}
\end{table}

\begin{table}
\centering
  \begin{minipage}{14cm}
    \renewcommand{\arraystretch}{1.2}
      \begin{tabular}{p{4cm}cccc}
        & $\sqrt{y_{ut}^2 + y_{tu}^2}$ & $\BR(t \to h u)$ 
        & $\sqrt{y_{ct}^2 + y_{tc}^2}$ & $\BR(t \to h c)$ \\ \hline

      \multicolumn{5}{l}{\bf Analysis~1: $th$ tag + top tag} \\
        \quad loose $th$ tags                    
        & $< 0.14$  &  $< 0.50 \%$  &  $< 0.14$  &  $< 0.53 \%$ \\
        \quad tight $th$ tags                    
        & $< 0.13$  &  $< 0.43 \%$  &  $< 0.13$  &  $< 0.48 \%$ \\

      \multicolumn{4}{l}{\bf Analysis~2: Higgs tag + top tag} \\
        \quad final cuts
        & $< 0.12$  &  $< 0.36 \%$  &  $< 0.24$  &  $< 1.5 \%$ \\
    \end{tabular}
  \end{minipage}
  \caption{Projected sensitivity of searches for anomalous $t \to j h$ decays and anomalous
  single top + Higgs production in the fully hadronic final state, using 100~fb$^{-1}$
  of 13~TeV data. See text for a detailed explanation of the different analyses.}
  \label{tab:hadronic-sensitivity}
\end{table}

\subsubsection{Analysis~2: Higgs tag + top tag}

For analysis~2, we identify events in which a Higgs boson is directly produced
(``Higgs tag'') by using the mass drop tagger implemented in
FastJet~3.0.3~\cite{Butterworth:2008iy, Cacciari:2011ma}. Following
\cite{Cacciari:2011ma}, we require the two subjets obtained when the last step
of clustering is undone to have jet masses at least a third smaller than the
mass of the original fat jet.  In addition, we require the asymmetry parameter
$y$~\cite{Cacciari:2011ma} to be larger than 0.09, thus making sure that both
subjets have a sizeable angular separation and each of them carries a
substantial fraction of the fat jet $p_T$.  If the latter is not the case for
one of the subjets, it is discarded and the algorithm is restarted with the
other subjet as input.  To remove contamination from pile-up and from the
underlying event (which we do not explicitly simulate), we filter the fat jet
by reclustering it with a smaller radius and keeping only the three hardest
constituents (see~Refs.~\cite{Butterworth:2008iy, Cacciari:2011ma} for details).
We require $b$ tags in the two hardest of them.

\begin{figure}
\centering
  \begin{tabular}{cc}
    \hspace*{-1.0cm}
    \includegraphics[width=0.48\textwidth]{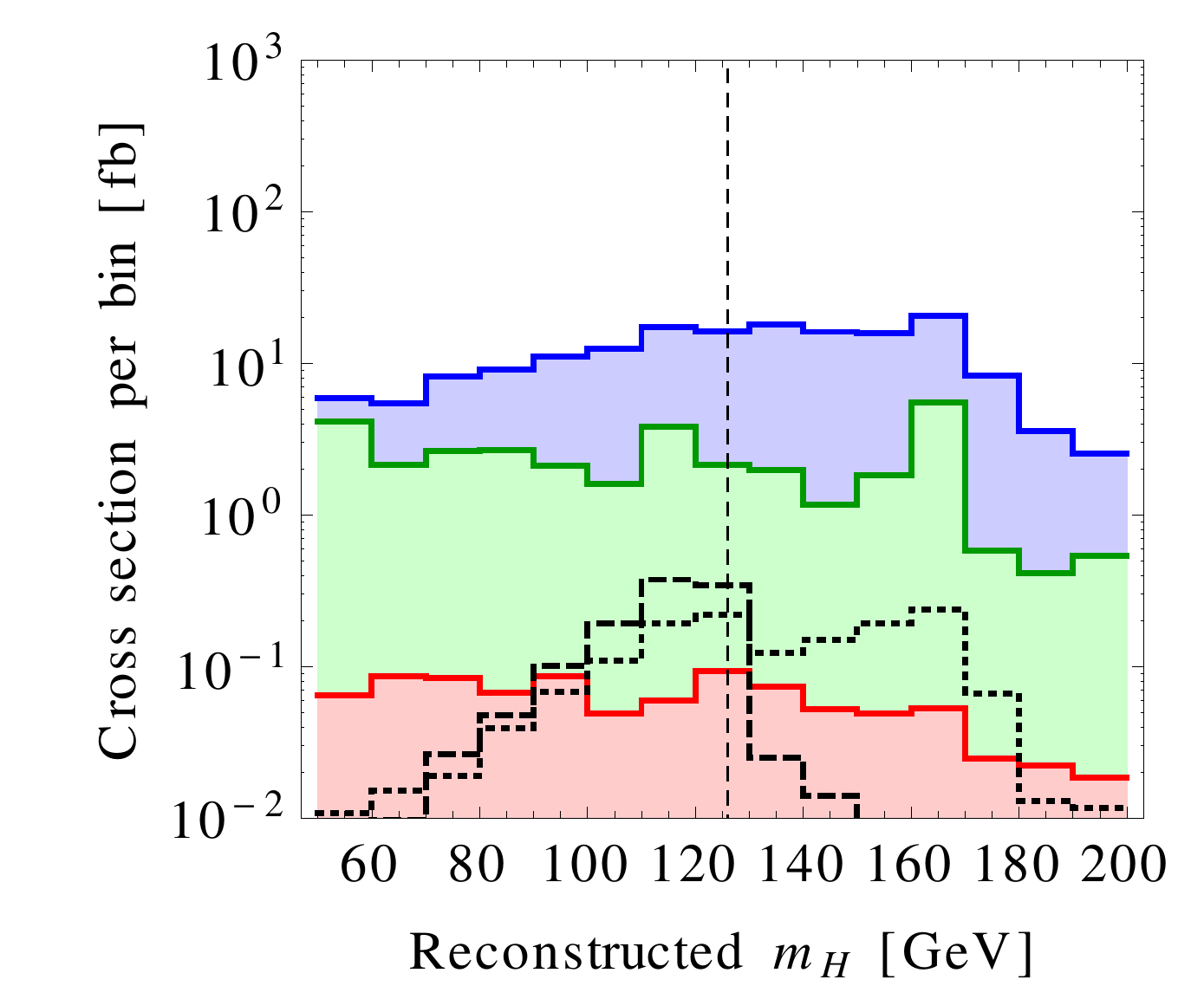} &
    \includegraphics[width=0.48\textwidth]{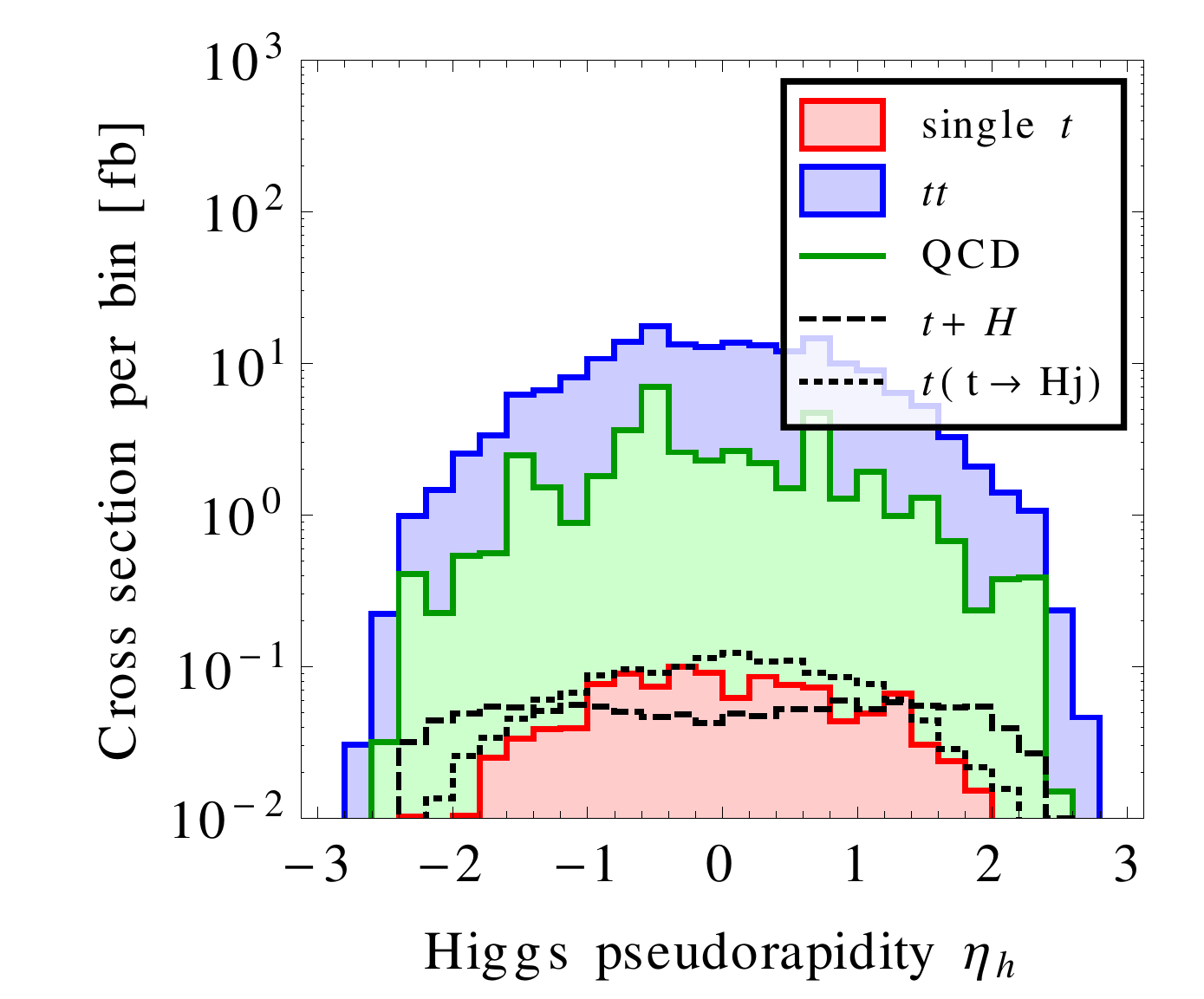} \\
    (a) & (b)
  \end{tabular}
  \caption{Kinematic distributions of events that pass the preselection
    of our search for $t+h$ production in the fully hadronic final
    state.  We show (a) the invariant mass $m_H$ and (b) the pseudorapidity $\eta_h$
    of the fast jet identified as the Higgs candidate.}
  \label{fig:toptag-kinematics}
\end{figure}

After top tagging and Higgs tagging, we preselect events by requiring that at
least one fat jet in the event carries a Higgs tag and at least one of the
remaining fat jets carries a top tag.  We define the Higgs candidate as the
hardest Higgs-tagged fat jet and the top candidate as the hardest top-tagged
fat jet different from the Higgs candidate.  If the hardest Higgs-tagged fat
jet is the only fat jet carrying a top tag, we take it to be the top candidate
and use the next-to-hardest Higgs-tagged fat jet as the Higgs candidate.  Event
counts after preselection are given in Table~\ref{tab:hadronic-event-rates},
and the distributions of two important kinematic quantities---the invariant mass
$m_H$ and the pseudorapidity $\eta_h$ of the Higgs candidate---are shown in
Fig.~\ref{fig:toptag-kinematics} for $\sqrt{y_{ut}^2 + y_{tu}^2} = 0.1$.
As expected, $m_H$ peaks around the true value of the Higgs mass for the signal,
while showing no distinct features for the background.
The forward bias of the $\eta_h$ distribution for
signal events is again related to angular momentum conservation in the center
of mass frame and the net boost of that frame in the direction of the incoming
up quark in the process $gu \to th$ (see Sec.~\ref{sec:setup} for details), making
$\eta_h$ again a good discriminant between $tuh$ and $tch$ couplings.
The $m_H$ and $\eta_h$ distributions shown in Fig.~\ref{fig:toptag-kinematics}
suggest the final cuts  $100\ \text{GeV} < m_H < 130\ \text{GeV}$ and $|\eta_h|
> 1.5$.  We see from the predicted event counts in the last row of
Table~\ref{tab:hadronic-event-rates} that the signal-to-background ratio $S/B$
is substantially improved by these cuts. Even though the signal-to-square root
background ratio $S/\sqrt{B}$ is similar before and after the final cuts, this
improvement makes the search much more robust with respect to systematic
uncertainties. 

From the event counts in Table~\ref{tab:hadronic-event-rates} and the projected
sensitivities in Table~\ref{tab:hadronic-sensitivity}, we see that analysis~2
outperforms analysis~1 in the case of $tuh$ couplings, but is not competitive
for $tch$ couplings, as expected.  It could therefore be an important
ingredient in a multi-channel search for $tuh$ couplings, and an important
cross check in case a signal is found in a different channel.

\section{Discussion and Conclusions}
\label{sec:conclusions}

In this work we have investigated the sensitivity of the LHC to flavor
violating top--Higgs interactions. Since these interactions are highly
suppressed in the SM, a positive signal at the LHC would constitute a clear
sign of new physics, for instance in the form of additional Higgs bosons or
nonrenormalizable couplings of the Higgs.

While exiting experimental searches have mainly concentrated on anomalous top
decays $t \to h q$, we have shown that anomalous single top plus Higgs
production is almost as important in the case of $tuh$ couplings and therefore
offers a promising avenue for further improvements in the sensitivity.  Single
top + Higgs production is less relevant for probing $tch$ interactions due to
the suppressed charm quark parton distribution in the proton. 

In Sec.~\ref{sec:improved-limits}, we have recast existing searches for
multilepton~\cite{CMS:2013jfa}, diphoton + lepton~\cite{CMS:2014qxa} and vector
boson + Higgs~\cite{CMS:ckv} final states to derive improved limits on $tuh$
couplings, including the contribution form single top + Higgs production.  Our
best limits on the branching ratio $\BR(t\to hu) < 0.45\%$ and the Yukawa
couplings $y_{ut}^2 + y_{tu}^2 < 0.014$ come from the diphoton plus leptons
final state and are a factor 1.5 stronger than the previously derived limits on
$\BR(t\to hc)$ and $y_{ct}^2 + y_{tc}^2$.  Limits from multileptons and vector
boson + Higgs searches are slightly weaker, but still competitive.
Our new limits are summarized in the upper part of Table~\ref{tab:sve}.

\begin{table}
\centering
  \begin{minipage}{16cm}
    \renewcommand{\arraystretch}{1.2}
      \begin{tabular}{p{6cm}cccc}
        & $\sqrt{y_{ut}^2 + y_{tu}^2}$ & $\BR(t \to h u)$ 
        & $\sqrt{y_{ct}^2 + y_{tc}^2}$ & $\BR(t \to h c)$ \\ \hline

      \multicolumn{5}{l}{\bf New limits from existing data} \\
        \quad Sec.~\ref{sec:multi}: Multilepton            
        & $< 0.19 $ & $< 1.0\%$ & $< 0.23 $ & $< 1.5\%$ \\
        \quad Sec.~\ref{sec:diphoton}: Diphoton plus lepton         
        & $< 0.12 $ & $< 0.45\%$ & $< 0.15 $ & $< 0.66\%$ \\
        \quad Sec.~\ref{sec:WH}: Vector boson plus Higgs         
        & $< 0.16 $ & $< 0.70\%$ & $< 0.21 $ & $< 1.2\%$ \\ \hline

      \multicolumn{4}{l}{\bf Projected future limits (13~TeV, 100~fb$^{-1}$)} \\
        \quad  Sec.~\ref{sec:WH}: Vector boson plus Higgs 
        & $< 0.076 $ & $< 0.15\%$ & $< 0.084 $ & $<  0.19\%$ \\
        \quad  Sec.~\ref{sec:multilepton-future}: Multilepton
        & $< 0.087 $ & $< 0.22\%$ & $< 0.11 $ & $<  0.33\%$ \\
        \quad Sec.~\ref{sec:hadronic}: Fully hadronic
        & $< 0.12 $ & $< 0.36\%$ & $< 0.13 $ & $<  0.48\%$ \\
    \end{tabular}
  \end{minipage}
  \caption{Summary of our new limits on flavor violating $tuh$ and $tch$ couplings from the
    CMS multilepton search, diphoton plus lepton search and vector boson plus
    Higgs search, as well as the projected sensitivities in a future multilepton
    search, a vector boson plus Higgs search and an analysis of fully hadronic
    final states using 100~fb$^{-1}$ of 13~TeV data. See text for a detailed
    explanation of the different analyses.}
  \label{tab:sve}
\end{table}

In the second part of the paper, Sec.~\ref{sec:future}, we have investigated
possible future improvements of searches for flavor violating top--Higgs couplings,
including the development of a completely new search strategy in fully hadronic
final states.  We have shown that multilepton, diphoton + lepton and vector
boson + Higgs searches can substantially improve the current bounds and may
have the potential to distinguish $tuh$ couplings from $tch$ couplings at the
$2\sigma$ level once a signal is discovered at $5\sigma$.  This is possible
because, in the case of $tuh$ couplings, the process $ug \to th$ contributes
significantly to the signal. In this process, the Higgs boson tends to be
produced with a large forward boost, while in all other signal processes the
Higgs rapidity distribution is more central.  Moreover, $ug \to th$ leads to an
asymmetry of the total charge of the final state leptons.  For $tch$ couplings,
the corresponding process $cg \to th$ is suppressed by the parton distribution
function of the charm quark and is therefore negligible.

Regarding the fully hadronic processes $(t \to bjj) + (h \to b\bar{b})$ and $t
\to j + (h \to b \bar{b})$, we have developed an analysis using jet
substructure techniques to tag SM top decays, $h \to b \bar{b}$ decays and $t
\to j + (h \to b \bar{b})$ decays.  We find that backgrounds can be suppressed
efficiently in such a search, leading to a sensitivity that is competitive to
that of searches with leptonic or semileptonic final states.  Our projected
future limits are summarized in the lower part of Table~\ref{tab:sve}.

For completeness we note that several other LHC processes exhibit potential
sensitivity to flavor violating top--Higgs interactions. For example, $tuh$
couplings can lead to an enhancement of di-Higgs production at tree level
through $u$--$u$ collisions with $t$-channel top exchange. However, in this
case the relevant cross-section scales with the fourth power of the flavor
violating Yukawa couplings, and at the current upper limit the resulting effect
is already subleading compared to the (already very suppressed) SM
rate~\cite{Baglio:2012np}.\footnote{Using FeynArts and
FormCalc~\cite{Hahn:2000jm}, we have also checked that possible $y_{tq,qt}$
loop contributions to gluon fusion induced di-Higgs
production~\cite{Djouadi:1999rca} are negligible given current constraints on
these couplings.} Similarly, same-sign top production from $u$--$u$ scattering
via Higgs exchange in the $t$-channel is expected to be below the current
experimental sensitivity (cf.~\cite{ATLAS:2012hpa}) given currently allowed
values of $y_{tu,ut}$.

To summarize, several signatures of flavor violating $tqh$ interactions at the
LHC which we have studied in the present paper exhibit comparable prospects to
constrain or discover such phenomena. Moreover, it may be possible to even
discriminate between $tuh$ and $tch$ signals by exploiting the presence of absence
of the partonic process $ug \to th$.  When multiple searches are combined into
a global analysis, they could allow the LHC experiments to probe the flavor
violating top--Higgs interactions well into the region of $\BR(t\to h j)
\lesssim 0.1\%$.

\section*{Acknowledgements}
We would like to thank Yan Wang for sharing the results on the NLO QCD
corrections to associated top plus Higgs production. We also thank Tilman
Plehn, Torben Schell and Peter Schichtel for very useful discussions. AG would
like to thank Luka Leskovec for his help with computing facilities used in this
work.  JK acknowledges the hospitality of CERN and of the Aspen Center for
Physics (supported by the US National Science Foundation under Grant
No.~1066293), where part of this work has been carried out.  This work was
supported in part  by the Slovenian Research Agency. 

\appendix

\section{Tagging Top Decays to Higgs + Jet}
\label{sec:th-tagger}

\begin{figure}
\centering
  \begin{tabular}{ccc}
    \includegraphics[width=0.32\textwidth]{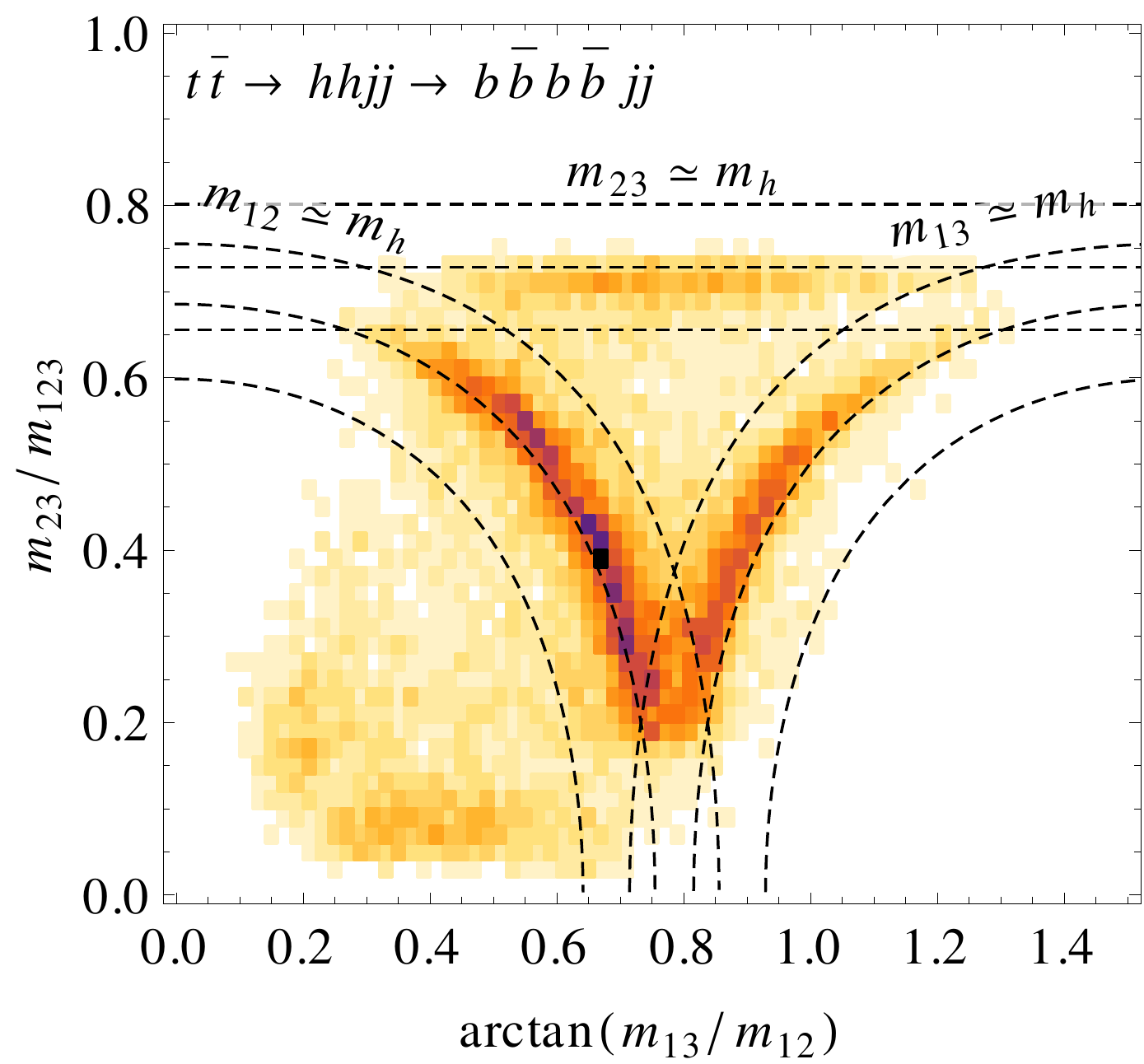} &
    \includegraphics[width=0.32\textwidth]{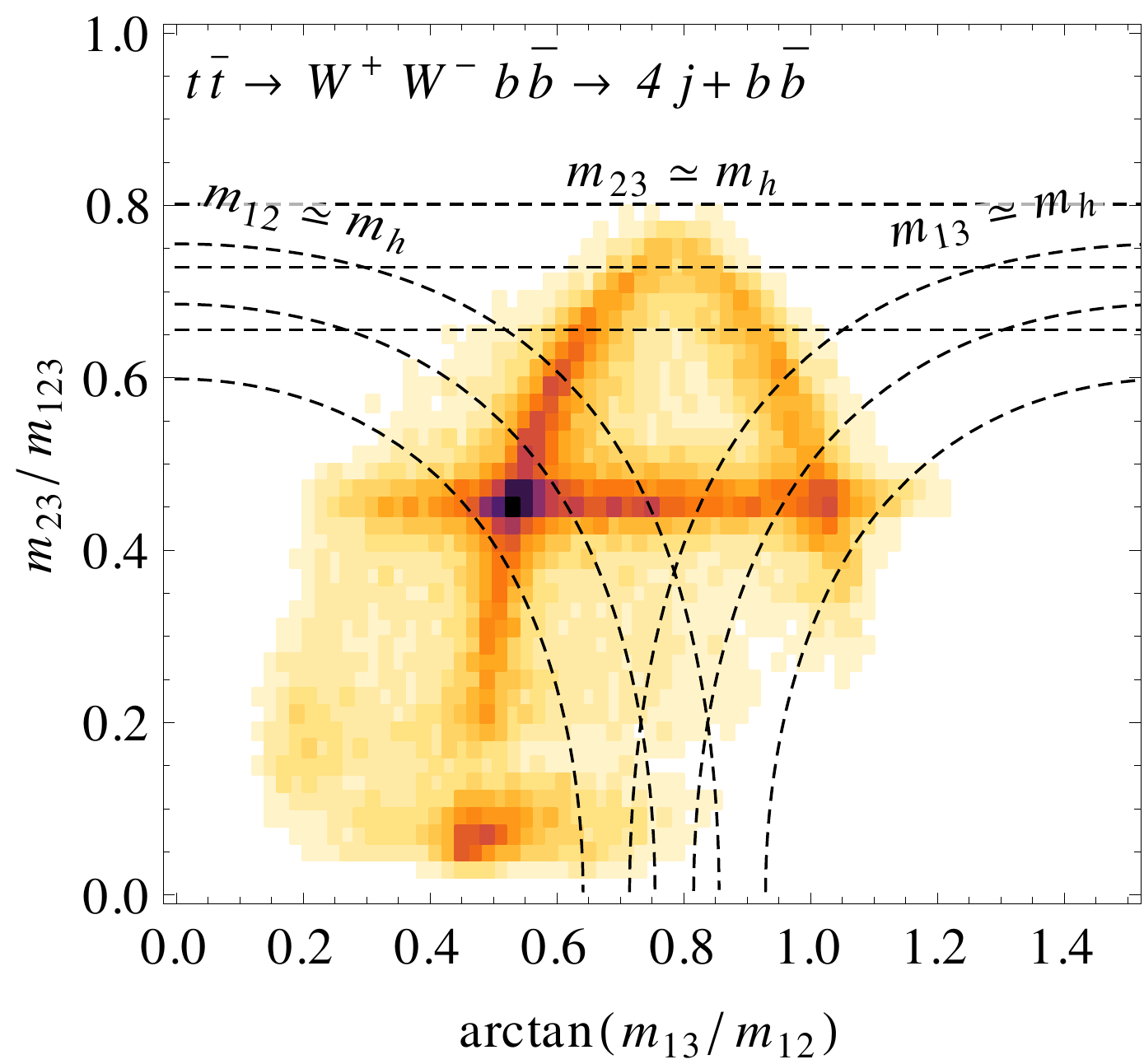} &
    \includegraphics[width=0.32\textwidth]{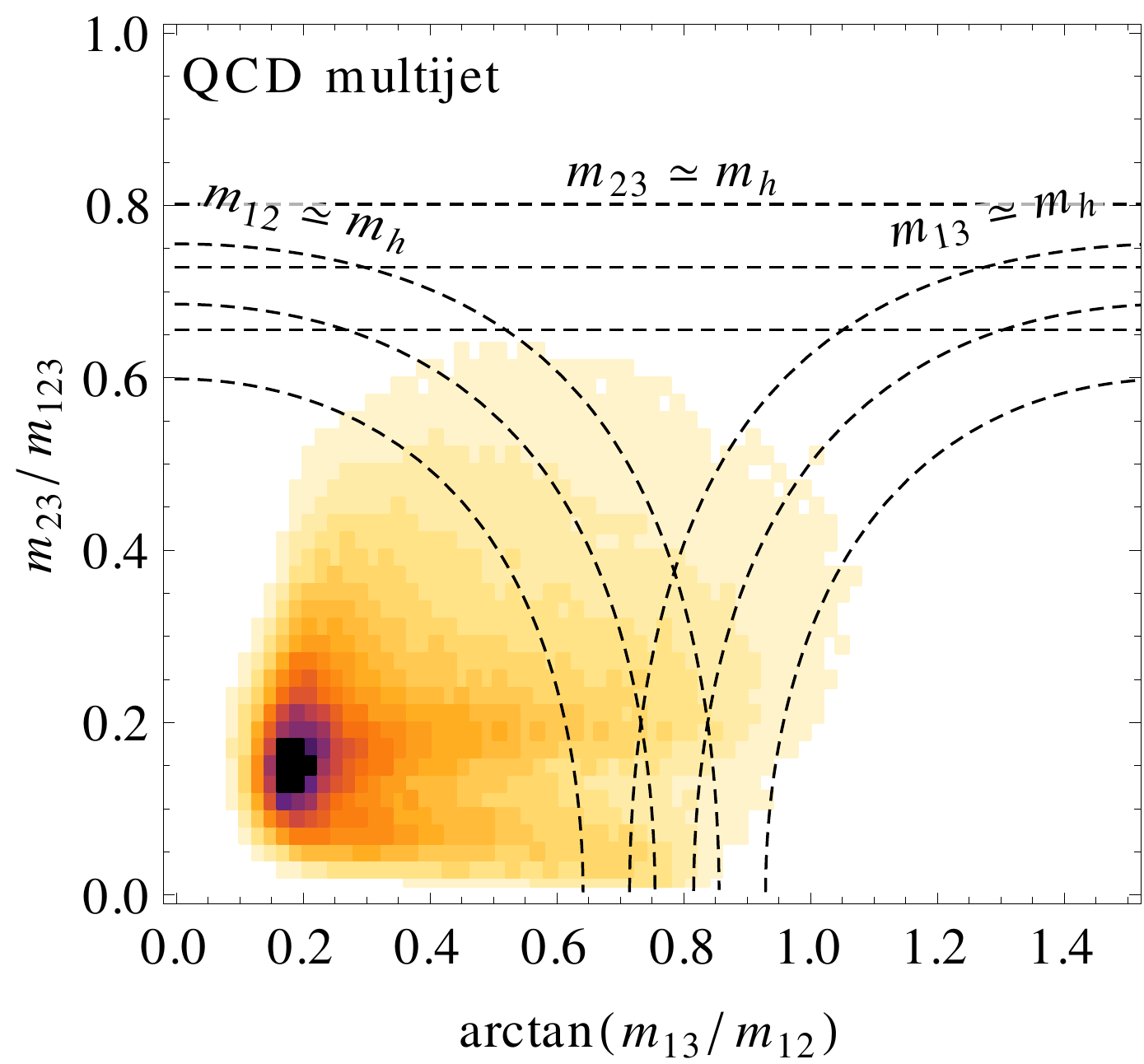} \\
        (a) & (b) & (c)
  \end{tabular}
  \caption{Subjet invariant mass distributions for (a) $t \bar{t} \to h h j j
    \to b \bar{b} b \bar{b} j j$ events, (b) $t \bar{t} \to b \bar{b} W^+ W^-
    \to b \bar{b} j j j j$ events, (c) QCD multijet events.  $m_{ij}$ denotes
    the invariant mass of the $i$-th and $j$-th HEPTopTagger subjets after
    filtering and reclustering into exactly 3 subjets.  Subjets are ordered by
    decreasing transverse momentum. The $R_\text{min}$ and $R_\text{max}$
    dependent cuts from Eq.~\eqref{eq:th-tagger-cuts} restrict the analysis to
    events lying within the dashed bands.  We have used $R_\text{min} = 0.9
    m_H / m_t$ and $R_\text{max} = 1.1 m_H / m_t$.  The median of the dashed
    bands corresponds to $m_{ij} = m_h$ for one combination of
    $i,\, j = 1,\, 2,\, 3$.}
  \label{fig:th-tagger}
\end{figure}

In this appendix, we give details on the ``$th$'' tagging algorithm used in
Sec.~\ref{sec:hadronic} to identify hadronic $t \to j + (h \to \bar{b} b)$
events.  Our method is based on the HEPTopTagger
algorithm~v1.0~\cite{Plehn:2009rk, Plehn:2010st}, a detailed description of
which is given in the Appendix of~\cite{Plehn:2010st}.  In simplified terms
HEPTopTagger starts from a fat jet, which it unclusters partially to identify
the three subjets that are most likely to originate from a top decay based on
their invariant mass $m_{123}$.  The algorithm then imposes cuts on the
invariant masses $m_{12}$, $m_{13}$ and $m_{23}$ of different pairings of these
three subjets, where the indices 1, 2 and 3 stand for the subjet with the
largest, next-to-largest and smallest $p_T$, respectively.  In particular, one
of the following three conditions has to be satisfied~\cite{Plehn:2010st}:
\begin{align}
  \begin{array}{r@{\quad}*3{>{\displaystyle}c}}
    \text{(i)} & 
    0.2 < \arctan \frac{m_{13}}{m_{12}}
      &\quad \text{and} \quad\qquad&
    R_\text{min} < \frac{m_{23}}{m_{123}} < R_\text{max}\,, \\[0.4cm]
    \text{(ii)} & 
    \frac{m_{23}}{m_{123}} > 0.35
      &\quad \text{and} \quad\qquad&
    R_\text{min}^2 \bigg[1 + \frac{m_{13}^2}{m_{12}^2}\bigg] < 1 - \frac{m_{23}^2}{m_{123}^2}
           < R_\text{max}^2 \bigg[1 + \frac{m_{13}^2}{m_{12}^2} \bigg]\,, \\[0.4cm]
    \text{(iii)} & 
    \frac{m_{23}}{m_{123}} > 0.35
      &\quad \text{and} \quad\qquad&
    R_\text{min}^2 \bigg[1 + \frac{m_{12}^2}{m_{13}^2}\bigg] < 1 - \frac{m_{23}^2}{m_{123}^2}
           < R_\text{max}^2 \bigg[1 + \frac{m_{12}^2}{m_{13}^2} \bigg]\,.
  \end{array}
  \label{eq:th-tagger-cuts}
\end{align}
The motivation for these cuts can be seen in Fig.~\ref{fig:th-tagger}, which
shows the distributions of $m_{23} / m_{123}$ and $\arctan(m_{13} / m_{12})$
for signal and background events.  The conditions on the left in
Eq.~\eqref{eq:th-tagger-cuts} loosely define the physically accessible region,
while the cuts on the right impose the condition that one of the subjet pairs
reconstructs to the mass of an on-shell intermediate particle: the $W$ for the
original HEPTopTagger and the Higgs for our $t \to h$ tagger.  Based on
Fig.~\ref{fig:th-tagger}, we choose $R_\text{min} = 0.9 \, m_H / m_t$,
$R_\text{max} = 1.1 \, m_H / m_t$ in our most conservative analysis (loose $th$
tags), and $R_\text{min} = 0.9 \, m_H / m_t$, $R_\text{max} = 1.0 \, m_H / m_t$ in
our more optimistic analysis (tight $th$ tags).\footnote{Note that a tight $th$
tag implies not only more restrictive cuts on $m_{23} / m_{123}$ and
$\arctan(m_{13} / m_{12})$, but also that the fat jet does not simultaneously
carry a top tag.} The latter is based on the observation that the invariant
masses of the subjets from Higgs decay tend to be slightly smaller than the
true $m_H$ on average. We attribute this to individual hadrons falling outside the fat
jet cone, being reconstructed as part of the wrong subjet, or being removed
by filtering.  Note that these effects are largest for the softest subjets.

As discussed in Sec.~\ref{sec:hadronic}, we require the two jets most likely
originating from a Higgs decay to contain $b$ tags.  In the absence of a full
detector simulation we perform $b$ tagging by searching for $b$ or $c$ quarks
with $p_T > 25$~GeV within an angular distance $\Delta R < 0.4$ from the
reconstructed subjet axis.  If a $b$ or $c$ quark satisfying these requirements
exists inside the subjet, we assign a $b$ tag with a probability depending on
the quark's transverse momentum $p_T$ and its pseudorapidity $\eta$ according
to
\begin{align}
  \text{for $b$ quarks:} \quad \epsilon_b^{(b)} &=
    \begin{cases}
      0.5 \tanh(0.03 \, p_T - 0.4) & \text{for $p_T > 15$~GeV and $|\eta| \leq 1.2$} \,, \\
      0.4 \tanh(0.03 \, p_T - 0.4) & \text{for $p_T > 15$~GeV and $1.2 < |\eta| \leq 2.5$} \,, \\
      0.0                          & \text{otherwise} \,,
    \end{cases} \\
  \text{for $c$ quarks:} \quad \epsilon_b^{(c)} &=
    \begin{cases}
      0.2 \tanh(0.03 \, p_T - 0.4) & \text{for $p_T > 15$~GeV and $|\eta| \leq 1.2$} \,, \\
      0.1 \tanh(0.03 \, p_T - 0.4) & \text{for $p_T > 15$~GeV and $1.2 < |\eta| \leq 2.5$} \,, \\
      0.0                          & \text{otherwise} \,.
    \end{cases}
\end{align}
If no sufficiently hard and central $b$ or $c$ quark is found, the probability
that the jet is still misidentified as a $b$ jet is
$\epsilon_b^{(\text{u,d,s})} = 0.001$. In practice, we do not actually discard
events, but merely reweight them with the appropriate tagging efficiency.

\bibliography{th}

\end{document}